\documentclass[showemail,showpacs,preprintnumbers,pre]{revtex4}

\usepackage{bm}
\begin{document}

\title{Absence of First-order Transition and Tri-critical Point in the
Dynamic Phase Diagram of a Spatially Extended Bistable System
in an Oscillating Field
}
\author{G.~Korniss$^{1}$}\email{korniss@rpi.edu}
\author{P.~A.\ Rikvold$^{2}$}\email{rikvold@csit.fsu.edu}
\author{M.~A.\ Novotny$^{3}$}\email{man40@ra.msstate.edu}
\affiliation{
$^1$Department of Physics, Applied Physics, and Astronomy, 
Rensselaer Polytechnic Institute, 
110 8th Street, Troy, NY 12180-3590 \\
$^2$Center for Materials Research and Technology, 
School of Computational Science and Information Technology, 
and Department of Physics,
Florida State University, Tallahassee, Florida 32306-4350 \\
$^3$Department of Physics and Astronomy, 
Mississippi State University, P.O. Drawer 5167,
Mississippi State, MS 39762-5167
}

\date{\today}

\begin{abstract}
It has been well established that spatially extended, bistable systems
that are driven by an oscillating field exhibit a 
nonequilibrium dynamic phase transition (DPT). The DPT occurs when the 
field frequency is on the order of the inverse of an intrinsic lifetime 
associated with the transitions between the two stable states in a
static field  
of the same magnitude as the amplitude of the oscillating field. The DPT 
is continuous and belongs to the same universality class as the 
equilibrium phase transition of the Ising model in zero field 
[G.~Korniss et al., Phys.\ Rev.\ E {\bf 63}, 016120 (2001);
H.~Fujisaka et al., Phys.\ Rev.\ E {\bf 63}, 036109 (2001)]. 
However, it has previously been claimed that 
the DPT becomes discontinuous at temperatures below a tricritical point 
[M.~Acharyya, Phys.\ Rev.\ E {\bf 59}, 218 (1999)]. 
This claim was based on observations in dynamic Monte Carlo simulations of 
a multipeaked probability density for the dynamic order parameter and negative 
values of the fourth-order cumulant ratio. Both phenomena can be
characteristic of discontinuous phase transitions. 
Here we use classical nucleation theory for the decay of metastable phases, 
together with data from large-scale dynamic Monte Carlo simulations of a 
two-dimensional kinetic Ising ferromagnet, to show 
that these observations in this case are merely finite-size effects. For 
sufficiently small systems and low temperatures, the continuous DPT is
replaced,  
not by a discontinuous phase transition, but by a crossover to 
stochastic resonance. In the infinite-system limit the stochastic-resonance 
regime vanishes, and the continuous DPT should persist  
for all nonzero temperatures. 
\end{abstract}

\maketitle

\section{Introduction}
\label{sec:I}

Metastability and hysteresis are exhibited by numerous natural and
artificial systems that are driven away from thermodynamic equilibrium 
by an external ``field." In the absence of such a field, 
and below some critical temperature or analogous control parameter, a large
class of such systems possess two equivalent, symmetry-broken
ordered phases. The external field selects one of these ordered phases as 
the global, stable minimum in the multidimensional 
free-energy landscape. The other ordered 
phase becomes metastable, separated from the basin of attraction 
of the stable phase by a 
free-energy barrier. If the system is
initially prepared in the metastable phase and thermal
fluctuations are present, the system eventually (possibly after an 
extremely long time) escapes from the
metastable free-energy well and approaches stable equilibrium. In the present
paper we consider the response of a {\em spatially extended\/},
bistable system driven by an external field which is periodic in time.
In particular, we focus on the {\em finite-size effects} of the
periodic system response.

Ferromagnets are perhaps the most commonly known systems 
that exhibit metastability
and hysteresis. In this paper we therefore use magnetic language, in
which the order parameter is the system magnetization $m$, and its
conjugate field is the external magnetic field $H$. Analogous
interpretations, e.g., using the terms ``polarization" and
``electric field" for ferroelectric systems, 
``coverage'' and ``electrochemical potential'' for adsorbate systems, etc., 
are straightforward \cite{SIDES99}. 

The dynamic response under an oscillating
external field can be viewed as a competition between two time
scales: the half-period $t_{1/2}$ of the external field 
(proportional to the inverse driving frequency) and the average
metastable lifetime $\langle\tau\rangle$ of the system (the mean
time spent in the metastable well) after a sudden field reversal. 
For low driving frequencies, the time-dependent 
magnetization oscillates about zero in synchrony with
the external field (symmetric dynamic phase). For high
frequencies, however, the magnetization does not have time to switch sign
during one half-period, and it oscillates about one
or the other of its degenerate zero-field values (asymmetric
dynamic phase). This symmetry breaking and the corresponding
dynamic phase transition (DPT) between the symmetric and the
asymmetric limit cycles of the system magnetization have attracted
considerable attention over the last decade. It was first observed
during numerical integration of the mean-field equation of
motion for the magnetization of a ferromagnet in an oscillating
field \cite{TOME90,Mendes91}. Since then it has been the focus of
investigation in numerous Monte Carlo simulations of kinetic Ising systems
\cite{LO90,SIDES96,SIDES97,SIDES98,SIDES99,ACHA95,ACHA97C,ACHA97D,ACHA98,%
ACHA99,BUEN00,KORN01},
further mean-field studies
\cite{Zimmer93,ACHA95,ACHA97D,ACHA98,BUEN98}, and most recently in
analytic studies of a bistable time-dependent Ginzburg-Landau (TDGL) model 
\cite{FUJI}. The DPT may also have been experimentally observed in Co
on Cu(001) 
ultrathin magnetic films \cite{HE,JIANG,SUEN} and recently in numerical studies
of fully frustrated Josephson-junction arrays \cite{JEON02} and 
anisotropic Heisenberg models \cite{HALL}. 
The results of these studies agree that there
exists a genuine continuous phase transition between the symmetric
and asymmetric dynamic phases, at least in some region of the
parameter space spanned by temperature, field, and half-period. 
Finite-size scaling studies of data from dynamic Monte Carlo simulations 
\cite{SIDES98,SIDES99,KORN01}, as well as analytic arguments \cite{FUJI} 
have demonstrated that this far-from-equilibrium phase transition belongs to 
the same universality class as the equilibrium Ising model in zero field. 
This result is consistent with previous symmetry \cite{GRIN85} and
renormalization group \cite{BASS94} arguments \cite{note1}. 

In their original paper on the DPT in a mean-field model \cite{TOME90}, 
Tom{\'e} and de~Oliveira reported that the continuous (second-order) 
phase transition observed at high temperatures appeared to 
change at a tricritical point (TCP) to a discontinuous 
(first-order) transition for low temperatures. Such a TCP was also reported in 
later mean-field work \cite{ACHA95}. However, an analytical and numerical 
mean-field study by Zimmer \cite{Zimmer93} makes a strong case that the claims 
for a TCP in the mean-field case is based on a misinterpretation of effects 
of critical slowing-down at the DPT. 
Similar claims, that in some
region of the dynamic phase diagram {\em spatially extended\/} 
kinetic Ising models exhibit a 
first-order transition and consequently have a TCP
separating lines of second-order and first-order dynamic phase
transitions, have also been made on the  
basis of dynamic Monte Carlo studies \cite{ACHA94,CHAK99,ACHA99}. 
For recent reviews on the DPT, see Refs.~\cite{CHAK99,ACHA94}.

The purpose of the present paper is to clear up the remaining 
confusion about the
interpretation of simulation results for the DPT in spatially extended 
kinetic Ising models, in particular
in the low-temperature regime where a first-order transition 
and a TCP have been claimed to exist \cite{ACHA94,CHAK99,ACHA99}. 
Those conclusions were essentially based on data for a single system size, 
and we here demonstrate how proper consideration of 
rather subtle finite-size effects 
leads to a different picture. The implication of our 
theoretical arguments and Monte Carlo simulations presented in this paper  
is that in an {\it infinitely large\/} system a continuous DPT 
should persist down to arbitrarily low temperatures. However, in any 
{\it finite\/} system for sufficiently low frequencies, the DPT gives
way to a transient regime of {\it stochastic resonance\/} (SR) \cite{SR} 
at a size-dependent temperature. It is 
this size-dependent crossover temperature, which has previously been 
misinterpreted as a TCP. 

The rest of this paper is organized as follows. 
In Sec.~\ref{sec:MSD} we summarize the theoretical
framework needed to understand the underlying metastable decay
mechanisms and their consequences for the DPT. This underscores
again the importance of the interplay of various time- and length
scales in metastable systems 
\cite{TOMI92,group1,SIDES96,SIDES97,SIDES99,SIDES98b}. 
In Sec.~\ref{sec:SIM} we extend our preliminary numerical work \cite{UGA01}, 
supporting our theoretical arguments by large-scale Monte
Carlo simulations of a two-dimensional 
kinetic Ising ferromagnet in an oscillating field. 
Our conclusions are summarized in Sec.~\ref{sec:CON}, 
and derivations of analytic approximations for quantities of
interest in the stochastic-resonance regime are given in
Appendix~\ref{sec:pq} and Appendix~\ref{sec:rtd}.

\section{Metastable decay modes and periodic response in finite systems}
\label{sec:MSD}

The appropriate dynamic order parameter in the presence of an
oscillating external field is the period-averaged
magnetization, $Q$$=$$\frac{1}{2t_{1/2}}\oint m(t)dt$ \cite{TOME90}. 
It takes a nonzero value in the asymmetric dynamic phase, while it 
vanishes in the symmetric phase. 
The transition occurs when the half-period
$t_{1/2}$ and the underlying metastable lifetime $\langle\tau(T,H)\rangle$
become comparable. The metastable lifetime 
depends on the temperature $T$ and the field amplitude $H$. For sufficiently
large systems (see quantitative statements below) the system
escapes from the metastable phase through the nucleation of many
droplets [multi-droplet (MD) regime \cite{TOMI92,group1}]. Consequently, the
time-dependent system magnetization is self-averaging. If
$\langle\tau(T,H)\rangle$$\ll$$t_{1/2}$, the magnetization follows
the external field in each half-period. The system relaxes to a
symmetric limit cycle, and the order-parameter probability density $P(Q)$ is
sharply peaked about $Q$$=$$0$. On the other hand, for
$\langle\tau(T,H)\rangle$$\gg$$t_{1/2}$ the magnetization does not have
enough time to switch within a single half-period, 
and it relaxes to an asymmetric limit cycle
(with occasional switches between the two equivalent asymmetric dynamic
phases). Consequently, $P(Q)$ becomes bimodal with sharp
peaks near $Q$$=$$\pm 1$. This breaking of the symmetry of the
limit cycle and the associated DPT have been carefully analyzed
\cite{SIDES99,SIDES98,KORN01} with finite-size scaling techniques
borrowed from equilibrium critical phenomena \cite{BIND92,FSS90}.
In terms of the dimensionless half-period,
$\Theta$$\equiv$$t_{1/2}/\langle\tau(T,H)\rangle$, the DPT occurs at a
critical value $\Theta_c$$\sim$${\cal O}(1)$. The finite-size scaling analysis 
of the Monte Carlo data also indicates that this far-from-equilibrium 
phase transition belongs to the same 
universality class as the {\em equilibrium\/} Ising model in zero field. 
Supporting these numerical studies, recent
analytic results within a coarse-grained TDGL model \cite{FUJI} indicate
that the behavior of the stochastic variable $Q$ is
governed by the effective Hamiltonian ${\cal H}_{\rm eff}$$=$$aQ^2
+ Q^4$, where $a$$\propto$$(\Theta - \Theta_c)$. 
According to standard arguments from the theory of critical phenomena, 
this result leads directly to the conclusion that the DPT belongs to the 
universality class of the zero-field Ising model in equilibrium \cite{GOLD92}, 
in agreement with the simulation results. 

For any {\em finite\/} system, however, the metastable decay mode
changes to the nucleation and growth of a {\em single} droplet
at sufficiently low temperatures 
[single-droplet (SD) regime \cite{TOMI92,group1}]. 
Due to the stochastic nature of the nucleation of a single droplet,
the corresponding response in the presence of an oscillating field
is different: the system exhibits \cite{SIDES97,SIDES98b,ACHA99} 
stochastic resonance \cite{SR}.

The crossover from the underlying  MD to SD decay can be
understood by using standard nucleation theory
applied to the ferromagnetic kinetic Ising model (in general dimension
$d$ and with ferromagnetic coupling constant $J$)
\cite{TOMI92,group1}.
{\em Below} the critical temperature, following a single, instantaneous 
field reversal, the average time between nucleation events
(the nucleation time) in a system of linear size $L$ is obtained as
\begin{equation}
t_{\rm n} = \left[L^{d}I(T,H)\right]^{-1} \;,
\label{nucl_time}
\end{equation}
where $I(T,H)$ is the temperature- and field-dependent nucleation
rate per unit volume. It can be expressed in terms of the free
energy $F(T,H)$ of the critical droplet as
\begin{equation}
I(T,H) = C(T,H)^{-1} e^{-F(T,H)/T} \;,
\label{nucl_rate}
\end{equation}
where $F(T,H)$ and the prefactor $C(T,H)$ can be obtained from
nucleation theory to various degrees of approximation, depending on
$T$ and $H$ \cite{group1}. The temperature $T$ is given in energy units 
by setting the Boltzmann constant $k_{\rm B}$$=$$1$ in Eq.~(\ref{nucl_rate}).
The other characteristic time scale is
the growth time $t_{\rm g}$. It is defined as the time it takes
for a supercritical droplet to grow to fill half the system
volume. Assuming a time-independent radial growth velocity $v(T,H)$, 
\begin{equation}
t_{\rm g}(L,T,H) = \frac{L}{[2\Omega_d(T)]^{1/d}v(T,H)}\;,
\label{growth_time}
\end{equation}
where $\Omega_d(T)$ is a dimension- and temperature-dependent shape
factor (between $\pi$ and 4 for $d$=2 and between $4 \pi /3$ and 8 for $d$=3) 
\cite{group1}. 

For $t_{\rm n}$$\ll$$t_{\rm g}$ 
{\em many} droplets nucleate while those nucleated
shortly after the field reversal are still growing. In this, the MD, regime
the lifetime is independent of the system size. An order-of-magnitude
estimate of the metastable lifetime $\langle\tau(T,H)\rangle$ can be obtained
\cite{group1} by
equating the volume $R_0^d$ reached by a droplet that grows for this amount of
time, $R_0^d$$\propto$$(v \langle \tau \rangle )^d$, 
with the volume inside which on
average a single nucleation event occurs during the same time,
$R_0^d$$\propto$$( \langle \tau \rangle I)^{-1}$ 
($R_0$ can be regarded as the typical droplet separation). The result is 
$\langle \tau \rangle \propto (v^d I)^{-1/(d+1)}$. 

For $t_{\rm n}$$\gg$$t_{\rm g}$ 
the {\em first} droplet to nucleate eventually fills the
system on its own. In this regime the lifetime depends strongly on the system
size and approximately equals $t_{\rm n}$ [Eq.~(\ref{nucl_time})]. 

\begin{figure}
\vspace*{2.0truecm}
\includegraphics{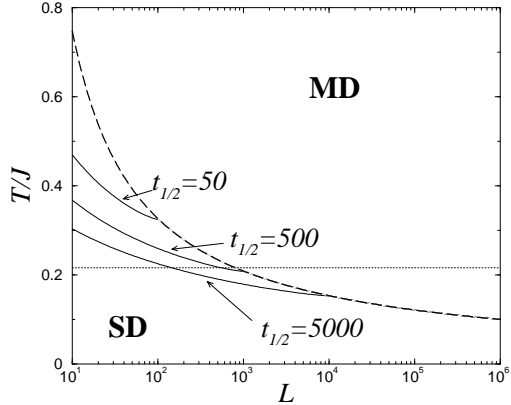}
\vspace*{4.0truecm}
\caption[]{Metastable decay modes and crossovers in the dynamic phase
diagram for the two-dimensional kinetic Ising ferromagnet in 
a square-wave oscillating field with amplitude $H$$=$$2.0J$. 
The dashed curve is the dynamic spinodal (DSP) $T_{\rm DSP}(L)$,
separating the underlying multi-droplet (MD) from the single-droplet
(SD) regime. The solid curves correspond to $T_{\times}(L)$
[Eq.~(\protect\ref{TX})], the crossover to the dynamic ``frozen''
state for various half-periods $t_{1/2}$, indicated by arrows [in
units of Monte Carlo steps per spin (MCSS)]. The horizontal dotted line
indicates the temperature $T$$=$$0.216J$, at which the dynamic phase
transition (DPT) occurs in the {\em large} system-size limit for
$t_{1/2}$$=$$500$ MCSS. Along this horizontal line, for $L$$\gtrsim$${\cal
O}(10^3)$ the metastable lifetime is system-size independent,
and its value is comparable to the fixed half-period.} 
\label{fig1}
\end{figure}

The above two decay modes characterize the MD and SD regime, respectively.
The crossover between the two regimes 
defines the dynamic spinodal (DSP) \cite{group1} and can
be estimated by equating $t_{\rm n}$ and $t_{\rm g}$. (In terms of the
underlying length scales, this corresponds to the situation in which
the mean droplet separation $R_0=v\langle \tau\rangle$ becomes comparable
to the system size $L$.) This yields an
implicit equation for the temperature corresponding to the DSP as
a function $L$ and $H$,
\begin{equation}
T_{\rm DSP}=\frac{F(T_{\rm DSP},H)}{(d+1)\ln L -\ln\{C(T_{\rm
DSP},H) [2\Omega_{d}(T_{\rm DSP})]^{1/d}v(T_{\rm DSP},H)\} } \;.
\label{T_DSP}
\end{equation}
For $d$$=$$2$, analytic approximations \cite{num_values} are known for $v(T,H)$
\cite{RIKOL00}, $\Omega_2(T)$, and $C(T,H)$ \cite{prefactor,BOMAN01}. The
corresponding estimate for $T_{\rm DSP}$ as a function of $L$ at $H=2.0J$,
obtained by numerically solving Eq.~(\ref{T_DSP}), is shown as a dashed
curve in Fig.~\ref{fig1}. The resulting curve $T_{\rm DSP}(L)$ turns
out to be quite insensitive for the precise values used in the
approximation \cite{num_values}. In the large-system limit $T_{\rm DSP}$
approaches zero logarithmically with increasing $L$. 
The existence of the DSP implies
that in the presence of an oscillating field, reducing $T$ at
fixed half-period $t_{1/2}$ and field amplitude $H$ results in
drastically different behaviors for ``small" and ``large" systems.

Upon reducing the temperature for sufficiently large $L$, the
$L$-independent metastable lifetime
$\langle\tau(T,H)\rangle$ becomes comparable to the fixed
half-period $t_{1/2}$ at a temperature above the one at which the 
underlying decay mode
would cross over to the SD mode. The horizontal line in Fig.~\ref{fig1}
corresponds to the temperature where
$\langle\tau(T,H)\rangle \approx t_{1/2} = 500$~MCSS. 
Near this line, the system exhibits a continuous DPT
\cite{SIDES99,SIDES98,KORN01}. 
Further lowering the temperature has no effect on the behavior of the
dynamic order parameter $Q$, since the system is already locked into one of its
symmetry-broken dynamic phases, where $P(Q)$ is sharply peaked
near $\pm 1$. From Fig.~\ref{fig1} it is clear that, in order to
observe the DPT in a finite system for a given value of $t_{1/2}$, 
one has to employ a sufficiently large system 
[e.g., $L$$\gtrsim$${\cal O}(10^3)$ for $t_{1/2}$=500~MCSS 
or $L$$\gtrsim$${\cal O}(10^2)$ for $t_{1/2}$=50~MCSS]. 
In general, the larger $t_{1/2}$ is, the larger systems one needs
in order to observe the $L$$\to$$\infty$ behavior.
This example illustrates that it is essential to keep in mind the finite-size
implications of the crossover between the MD and SD decay mechanisms.

Upon reducing the temperature for smaller $L$, 
the crossover from the MD to the SD regime
occurs at a temperature well above that at which the infinite-system 
DPT would occur. In the SD regime the switching is stochastic
\cite{group1}. The system will then exhibit stochastic resonance,
and the order-parameter distribution
$P(Q)$ will indeed possess multiple peaks, leading to a
negative fourth-order cumulant \cite{ACHA99}. Lowering
the temperature still further reduces the probability that the magnetization 
switches during a finite number of periods, and the system
becomes effectively ``frozen." The crossover to this frozen phase is 
approximately marked by the curve along which the probability is 1/2 
that the magnetization does not
switch during a half-period in which it is opposite to the direction of 
the field. This crossover curve can be
estimated by noting that SD switching is brought about by the nucleation of a
single droplet, which is a Poisson process. Thus, the probability
that the magnetization does not change sign during a half-period
in which it started off opposite to the field is
\begin{equation}
P_{\rm not}(t_{1/2};L,T,H)= \left\{
\begin{array}{lll}
\exp\left\{ {- \left[t_{1/2}-t_{\rm g}(L,T,H) \right] /t_{\rm n}(L,T,H)} \right\} 
& \mbox{for} & t_{\rm g} \leq t_{1/2} \\
1 & \mbox{for} & t_{\rm g} > t_{1/2}
\end{array}
\right.
.
\label{P_not}
\end{equation}
Setting $P_{\rm not} = 1/2$ leads again to an implicit
equation for the corresponding crossover temperature, 
\begin{equation}
T_{\times}=\frac{F(T_{\times},H)}{ d \ln L
+\ln\left[\frac{1}{C(T_{\times},H)\ln 2}\left(
t_{1/2}-\frac{L}{[2\Omega_{d}(T_{\times})]^{1/d}v(T_{\times},H)}
\right)\right]} \;.
\label{TX}
\end{equation}
Estimates of $T_{\times}$ at $H$$=$$2.0J$ for three different values of 
$t_{1/2}$, obtained by numerically solving Eq.~(\ref{TX})
\cite{num_values}, are shown in Fig.~\ref{fig1} as solid curves. 
Since $P_{\rm not}$ changes rapidly between zero and unity as $T$ is
reduced, the specific cut-off value of $P_{\rm not} = 1/2$, used to
define $T_{\times}$ here, is not essential. For each value of $t_{1/2}$, the
curves representing $T_{\rm DSP}$ and $T_{\times}$ form the
border of a wedge-shaped region in which SR is observed (see Fig.~\ref{fig1}). 
This is the regime where the dynamic order parameter is
indeed characterized by a multi-peaked probability density and a negative
fourth-order cumulant \cite{ACHA99}. However, the above theoretical
arguments imply that the SR behavior is a
finite-size effect. 
We therefore conclude 
that the otherwise sound simulation results that Acharyya obtained in the
stochastic regime \cite{ACHA99}, were misinterpreted by him as signs of a 
first-order DPT. 

To summarize the periodic response for the ferromagnetic Ising model,
for an infinite (or sufficiently large) system, the system undergoes a 
continuous DPT when the the half-period becomes comparable with the metastable
lifetime. Crossing the dynamic phase boundary (Fig.~\ref{fig1}) 
by {\em changing the temperature},
for fixed and low frequencies (long half-periods), the metastable lifetime
becomes comparable to the half-period at appropriately low
temperatures, thus the DPT occurs at a low temperature (horizontal
dotted line in Fig.~\ref{fig1}). 
For finite and too small systems,
however, the underlying decay mode crosses over to the SD regime {\em before}
the DPT occurs (crossing the dashed curve from above in Fig.~\ref{fig1}). 
Then the system exhibits SR in the wedge-shape region, until practically no
switching occurs during any finite observation time (frozen state).
As we show next, careful analysis of simulation
results for systems of different sizes reveals that the signatures 
attributed to a first-order DPT in Ref.~\cite{ACHA99} indeed
disappear in the $L$$\to$$\infty$ limit, as predicted by the theoretical 
arguments presented above.

\section{Simulation results and finite-size effects}
\label{sec:SIM}

To model spatially extended bistable systems we performed dynamic 
Monte Carlo simulations of a
two-dimensional  kinetic Ising ferromagnet below its equilibrium critical
temperature. This simple model has, for example, been shown to be
appropriate for describing magnetization dynamics in highly
anisotropic single-domain nanoparticles and uniaxial thin films
\cite{JIANG,group2,HE,SUEN}. 
Despite its simplicity, it is believed to capture the
generic features of periodically driven, spatially extended
bistable systems. The system, which is defined on a two-dimensional 
square lattice of linear size $L$, is described by the Hamiltonian
\begin{equation}
{\cal H} = -J\sum_{\langle i,j\rangle} s_i s_j
-H(t)\sum_{i=1}^{L^2}s_i \;,
\label{ising_hamil}
\end{equation}
where $s_i$$=$$\pm 1$ is the state of the $i$th spin, $J$$>$$0$ is
the ferromagnetic coupling constant, $\sum_{\langle i,j\rangle}$
runs over all nearest-neighbor pairs, and $H(t)$ is an oscillating,
spatially uniform applied field. We use a square-wave field with
amplitude $H$. This has obvious computational advantages over a
sinusoidal field, while it does not change the universal 
characteristics of the system response \cite{KORN01}. The dynamic used is the
single-spin-flip Glauber algorithm with updates at randomly chosen
sites \cite{BIND92,GLAUB63}. At temperature $T$, each attempted
spin flip from $s_i$ to $-s_i$ is accepted with probability
\begin{equation}
W(s_i \to -s_i)= \frac{e^{-\Delta E_i /T}}{1+e^{-\Delta E_i /T}}
\;,
\label{glauber_rates}
\end{equation}
where $\Delta E_i$ is the energy change that would result from the
accepted flip. We give the temperature $T$ in energy units by setting
the Boltzmann constant $k_{\rm B}$$=$$1$ in Eq.~(\ref{glauber_rates}). 
For the largest system sizes ($L$$\geq$$1024$) we
implemented a scalable massively parallel version of this dynamic
\cite{KORN99,KORN00}, first proposed by Lubachevsky \cite{LUBA}.

The dynamic order parameter \cite{TOME90} is the period-averaged
magnetization
\begin{equation}
Q = \frac{1}{2t_{1/2}}\oint m(t) dt \;,
\label{dyn_op}
\end{equation}
where $m(t)= L^{-2} \sum{_i}s_i(t)$. The beginning of the period is chosen at a
time when $H(t)$ changes sign. In particular, we compute {\em both}
types of period averages, staring at the instant when $H(t)$ changes
from $+H$ to $-H$, and also starting when it changes from $-H$ to
$+H$. Both observations are included in the order-parameter
histograms with equal weight. This simplest form of phase
averaging is sufficient to produce a symmetric distribution for $Q$,
in particular in the stochastic regime.

Large-scale simulations and finite-size scaling studies of the
DPT have been recently performed with both sinusoidal
\cite{SIDES99,SIDES98} and square-wave \cite{KORN01} fields. The
results imply that the system undergoes a continuous phase
transition as the half-period $t_{1/2}$ becomes comparable to the average
metastable lifetime $\langle \tau(T,H) \rangle$.
Recall that the lifetime becomes independent of the system size for large
systems. The critical exponents for the dynamic order parameter and its
fluctuations at the DPT are consistent with those of the
two-dimensional equilibrium Ising transition
\cite{SIDES99,SIDES98,KORN01,FUJI}. In those studies, the temperature
and the field amplitude were held fixed, resulting in a fixed
lifetime $\langle \tau(T,H) \rangle$. The DPT was approached by tuning the
half-period $t_{1/2}$ of the oscillating field so that it became
comparable to $\langle \tau(T,H) \rangle$. An advantage of this approach
is that if the smallest system is already in the MD regime,
all the larger ones are, as well. Thus, one does not have to deal
with subtle crossovers corresponding to the different underlying
decay modes (SD vs MD). In the present study, we keep $t_{1/2}$ fixed and
tune the metastable lifetime by varying the temperature
$T$. The motivation for this is to closely parallel the study by Acharyya
\cite{ACHA99}, and to show that ignoring the
finite-size effects and the resulting crossovers, one can
easily misinterpret the stochastic-resonance behavior in the
stochastic regime as indicating a first-order transition.

\begin{figure}
\vspace*{2.5truecm}
\includegraphics{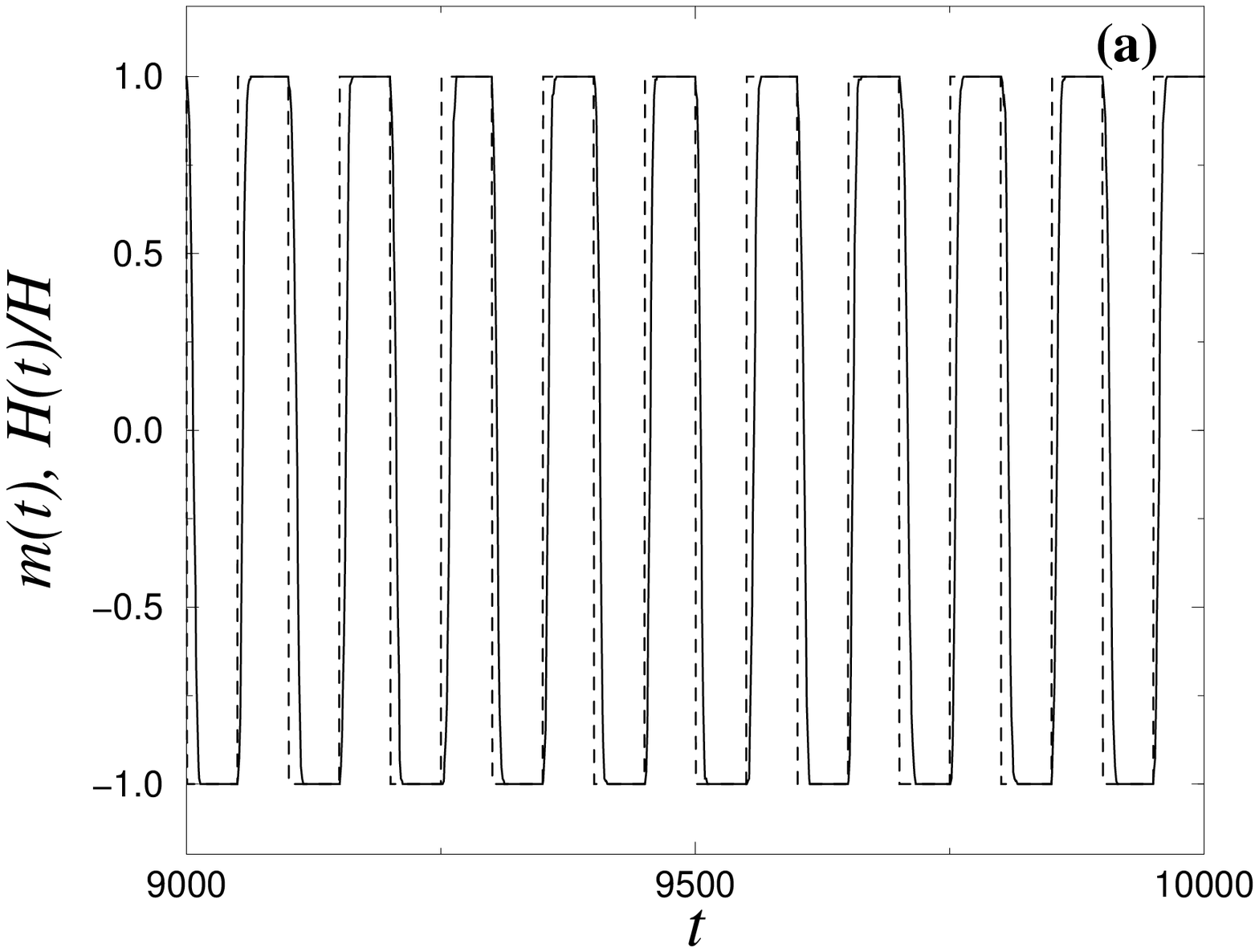}
\includegraphics{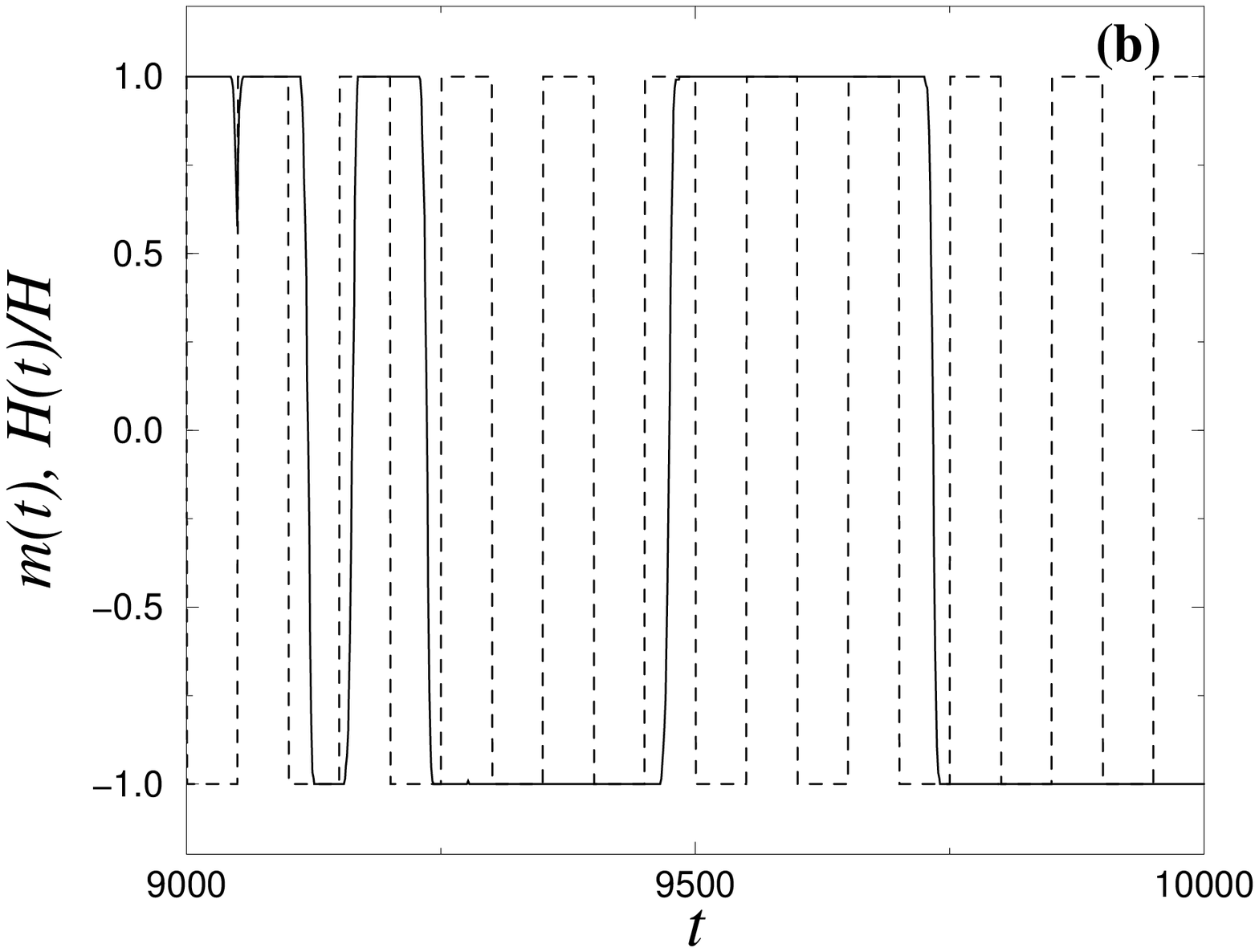}
\includegraphics{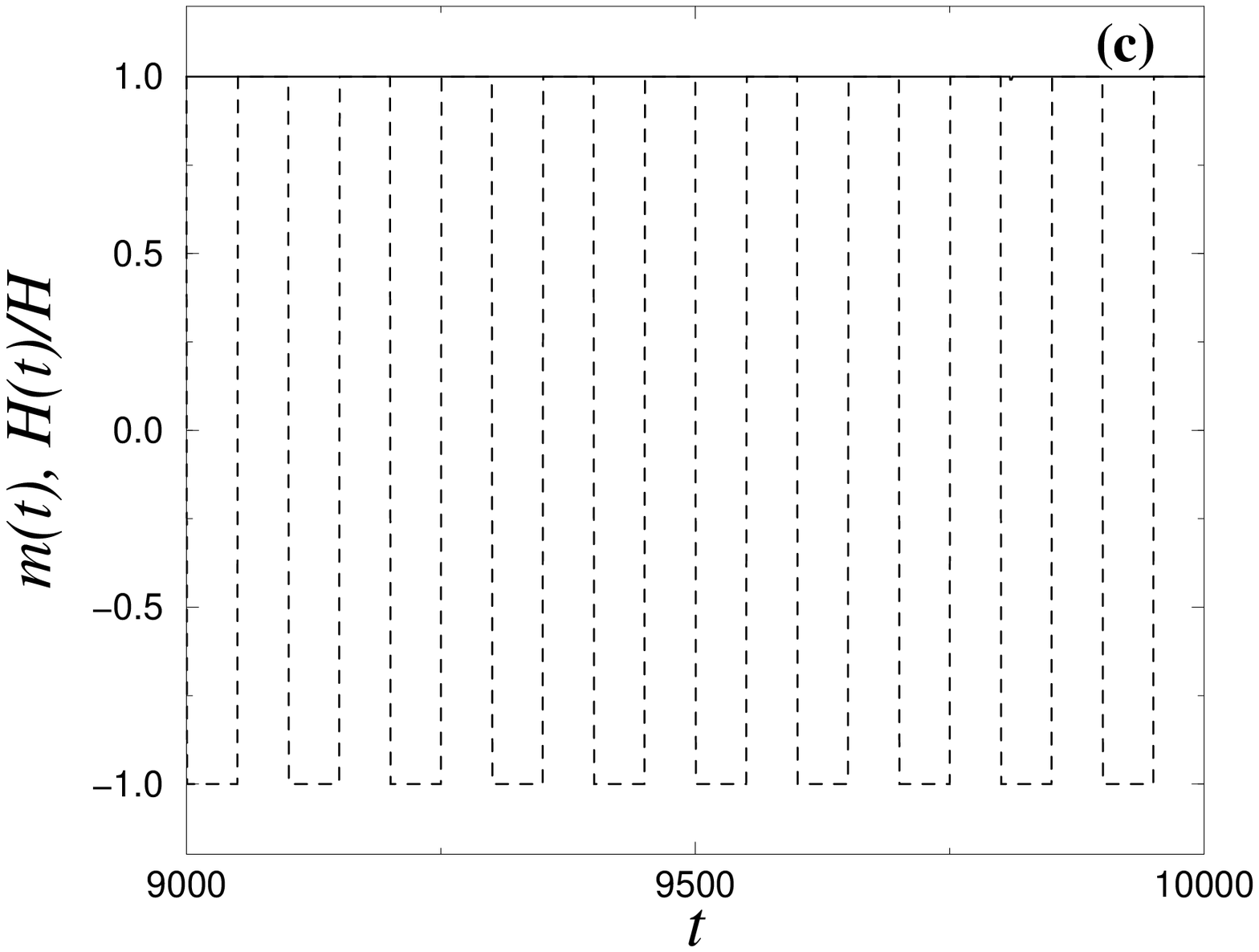}
\vspace*{2.5truecm}
\caption[]{Magnetization time series $m(t)$ (solid curves) and normalized 
applied field  $H(t)/H$ (dashed curves) with $H$$=$$2.0J$ and 
$t_{1/2} = 50$~MCSS, shown for a ``small'' system with $L$=16 
at different temperatures. 
(a) $T$$=$$0.8J$, dynamically disordered phase. 
(b) $T$$=$$0.4J$, stochastic resonance. 
(c) $T$$=$$0.35J$, dynamically ``frozen'' state.}
\label{fig2}
\end{figure}

\begin{figure}
\vspace*{2.5truecm}
\includegraphics{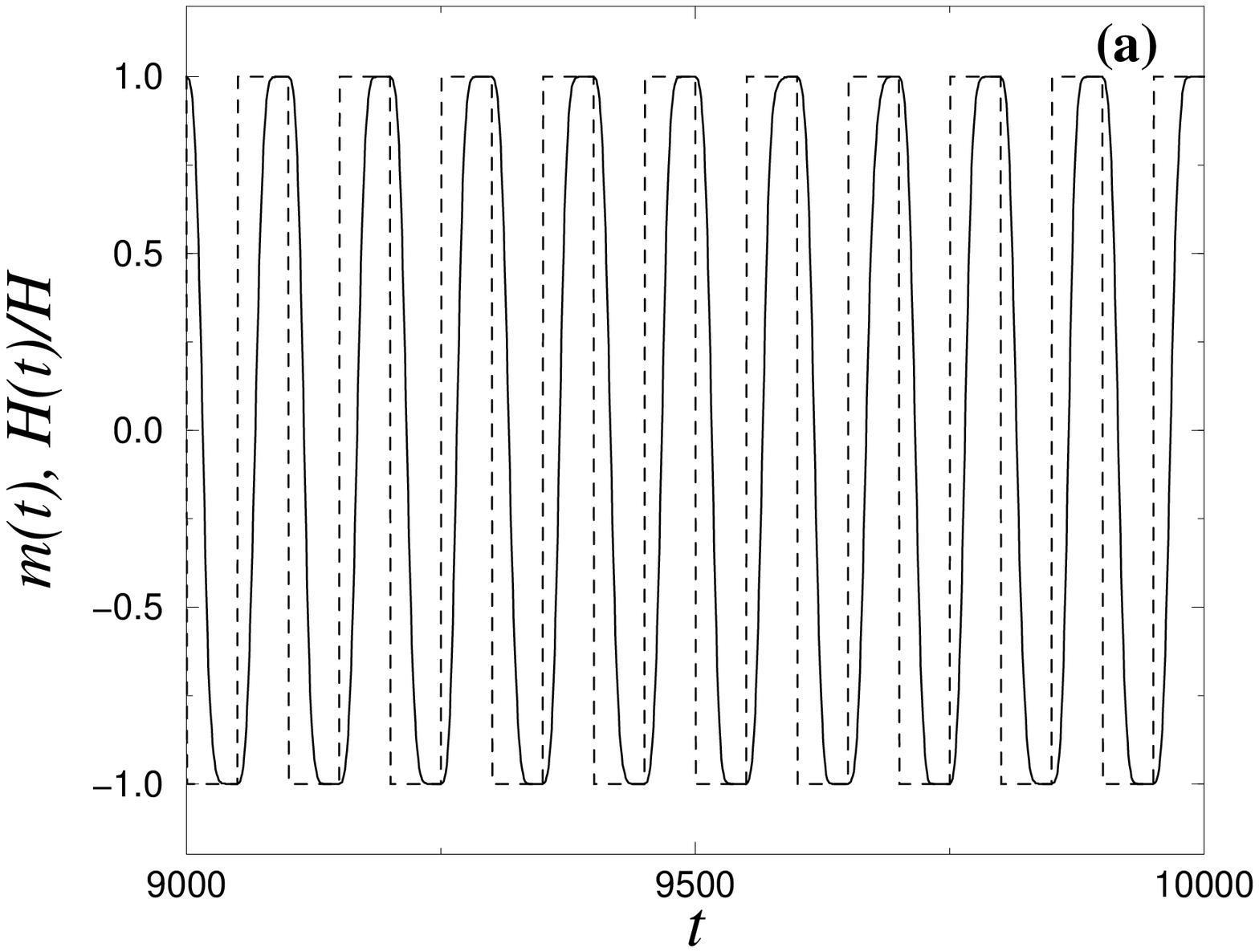}
\includegraphics{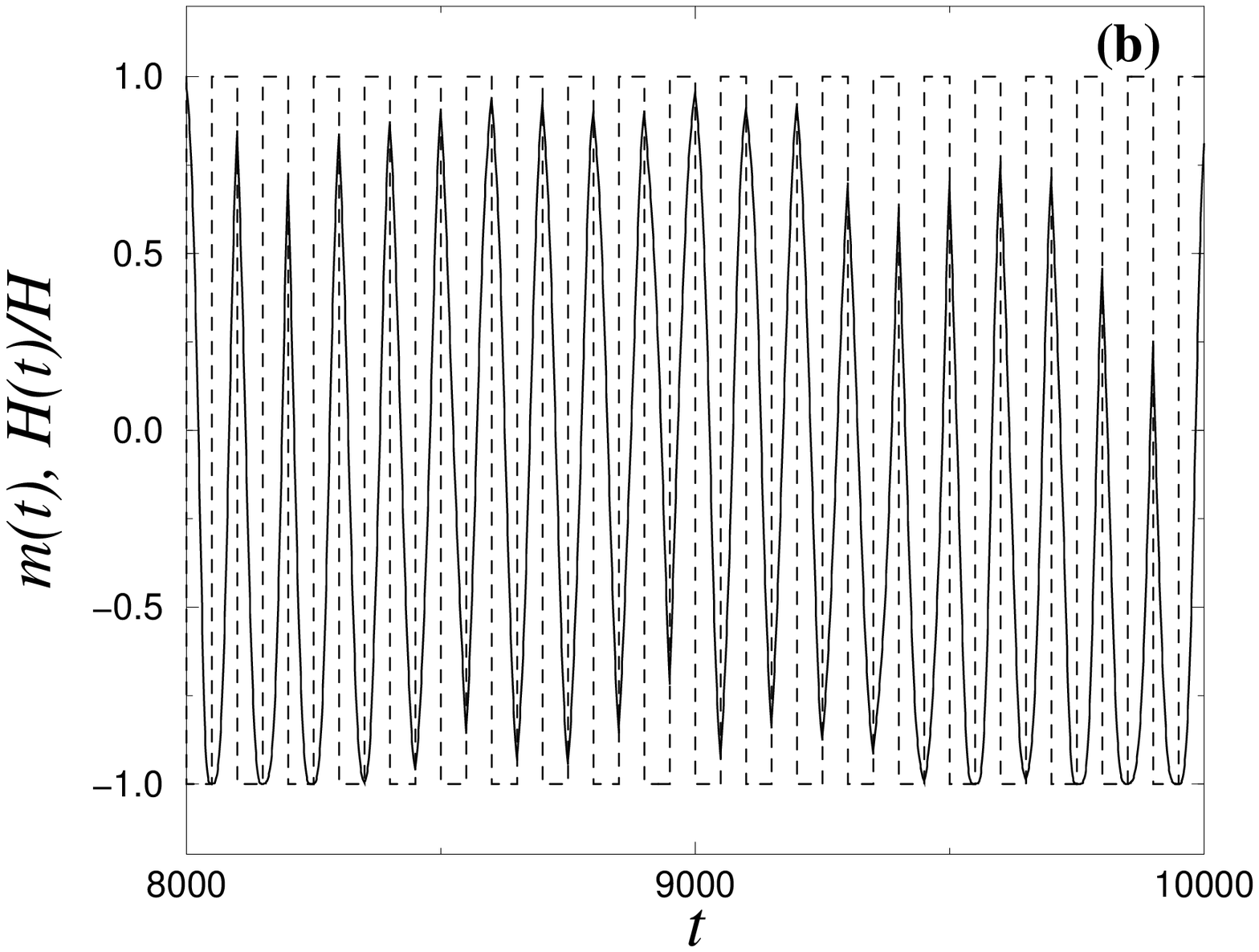}
\includegraphics{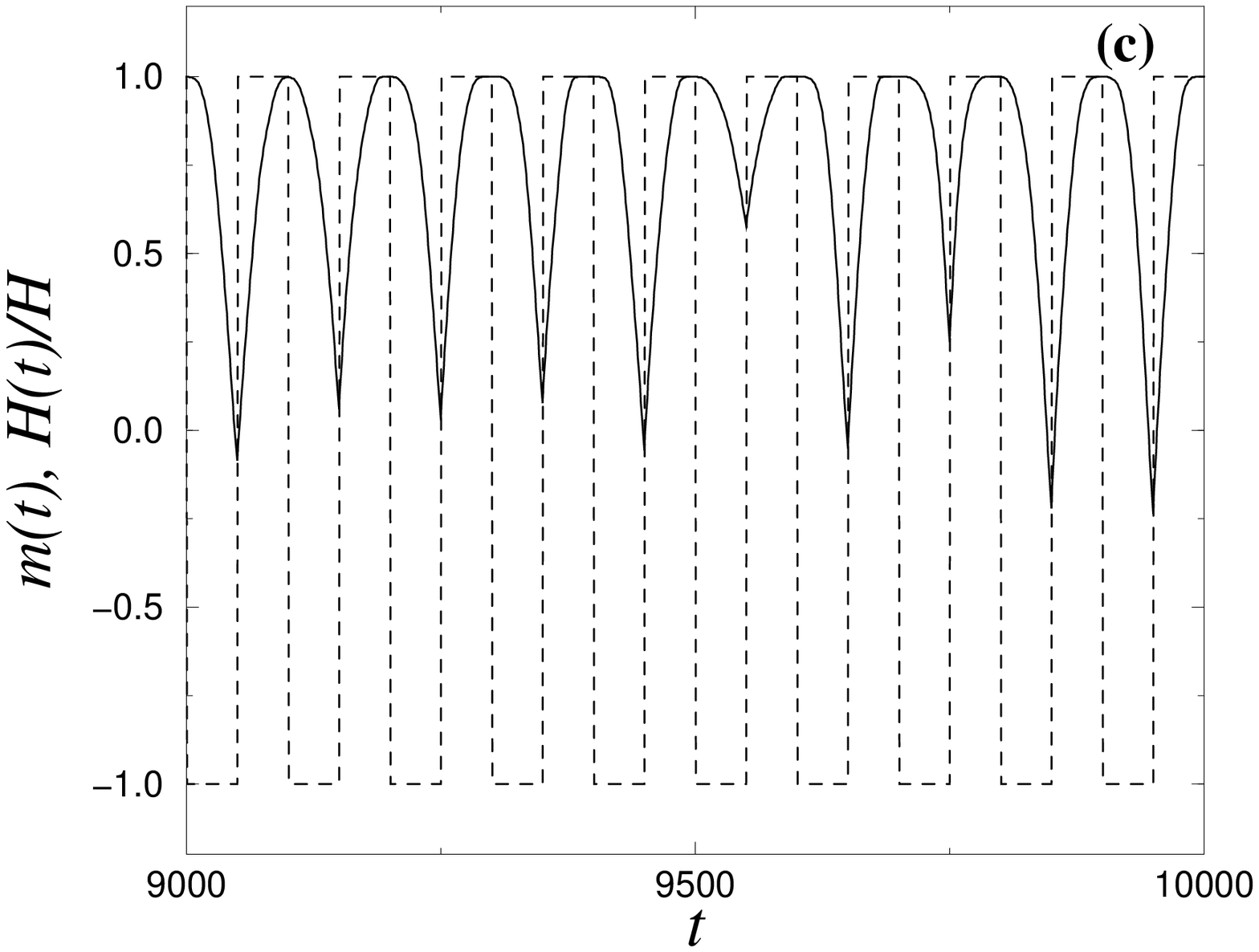}
\vspace*{2.5truecm}
\caption[]{Magnetization time series $m(t)$ (solid curves) and normalized 
applied field  $H(t)/H$ (dashed curves) with $H$$=$$2.0J$ and 
$t_{1/2} = 50$~MCSS, shown for a ``large'' system with $L$=180 
at different temperatures. 
(a) $T$$=$$0.5J$, dynamically disordered phase. 
(b) $T$$=$$0.375J$ near the DPT. 
(c) $T$$=$$0.34J$ dynamically ordered phase.
}
\label{fig3}
\end{figure}

Tracing the magnetization time series $m(t)$ already reflects the
major qualitative differences between the responses for ``small'' and
``large'' systems, as shown in Figs.~\ref{fig2} and \ref{fig3}, respectively. 
For sufficiently small $L$, as the temperature is reduced, the system
enters the stochastic regime, characterized by occasional random
switches [Fig.~\ref{fig2}(b)], before becoming completely
``frozen'' [Fig.~\ref{fig2}(c)]. 
For large $L$, the system undergoes a DPT
characterized by the slow wandering of the period-averaged
magnetization [Fig.~\ref{fig3}(b)] on its way to performing an
asymmetric limit cycle in the dynamically ordered phase 
at still lower temperatures [Fig.~\ref{fig3}(c)]. 

We performed simulations on system sizes ranging from $L$$=$$16$
to $2048$, choosing various field amplitudes $H$ and half-periods
$t_{1/2}$ that were kept fixed while the temperature $T$ was
varied. The time unit used is one Monte Carlo step per spin
[MCSS], i.e., one random ``sweep" of the $L$$\times$$L$ lattice. 
The lengths of the runs were $10^3$ full periods of the oscillating field for
$t_{1/2}$$=500$ MCSS and $10^4$ full periods for all the other 
values of $t_{1/2}$. Measuring the period-averaged
magnetization $Q$ after each half-period, we constructed averages
of the norm of the order parameter, $\langle|Q|\rangle_{L}$. We further 
calculated the fourth-order cumulant ratio,
\begin{equation}
U_{L} = 1 - \frac{\langle Q^4\rangle_L}{3\langle Q^2\rangle^2_L}
\;,
\label{cumulant}
\end{equation}
which typically provides a strong indication of the nature of any
underlying phase transition \cite{BIND92}. For a
continuous transition, $U_L$ changes monotonically from $0$ to
$2/3$ as one tunes the system from the symmetric (disordered) to
the symmetry-broken (ordered) phase. On the other hand, for a
first-order transition, $U_L$ develops a minimum, whose location
corresponds to the transition point. We also
constructed histograms of $Q$, representing the order-parameter
distribution $P(Q)$. In the stochastic regime, we in addition 
measured the residence times $t_{\rm r}$ and constructed their probability 
distribution $P_{\rm r}(t_{\rm r})$, the {\em residence-time distribution\/} 
(rtd). Here $t_{\rm r}$ is defined as the time spent
in one of the two ``wells" of the underlying system free energy
between two consecutive switching events. It is measured as the
time elapsed between two consecutive zero-crossings of
$m(t)$. The residence times and their distribution are
useful to characterize the system in the stochastic SD regime.


Keeping $H$ and $t_{1/2}$
fixed, we performed simulations measuring $\langle |Q| \rangle$ and $U_L$ as 
functions of $T$ for a series of system sizes. 
Here we present the results
for four different pairs of values of $H$ and $t_{1/2}$.
Figures~\ref{fig4}, \ref{fig5}, \ref{fig6}, and \ref{fig7} show
the results for $H$$=$$2.0J$ and $t_{1/2}$=50~MCSS, $H$$=$$2.0J$ and
$t_{1/2}$=500~MCSS, $H$$=$$1.8J$ and $t_{1/2}$=20~MCSS, and for
$H$$=$$3.0J$ and $t_{1/2}$=20~MCSS, respectively. 
As indicated by these figures, the findings in all these cases 
are qualitatively the same. 
For the purpose of discussion, we use Fig.~\ref{fig4}, corresponding to
$H$$=$$2.0J$ and $t_{1/2}$=50~MCSS.

\subsection{Small systems}
\label{sec:small}
\begin{figure}
\vspace*{2.5truecm} 
\includegraphics{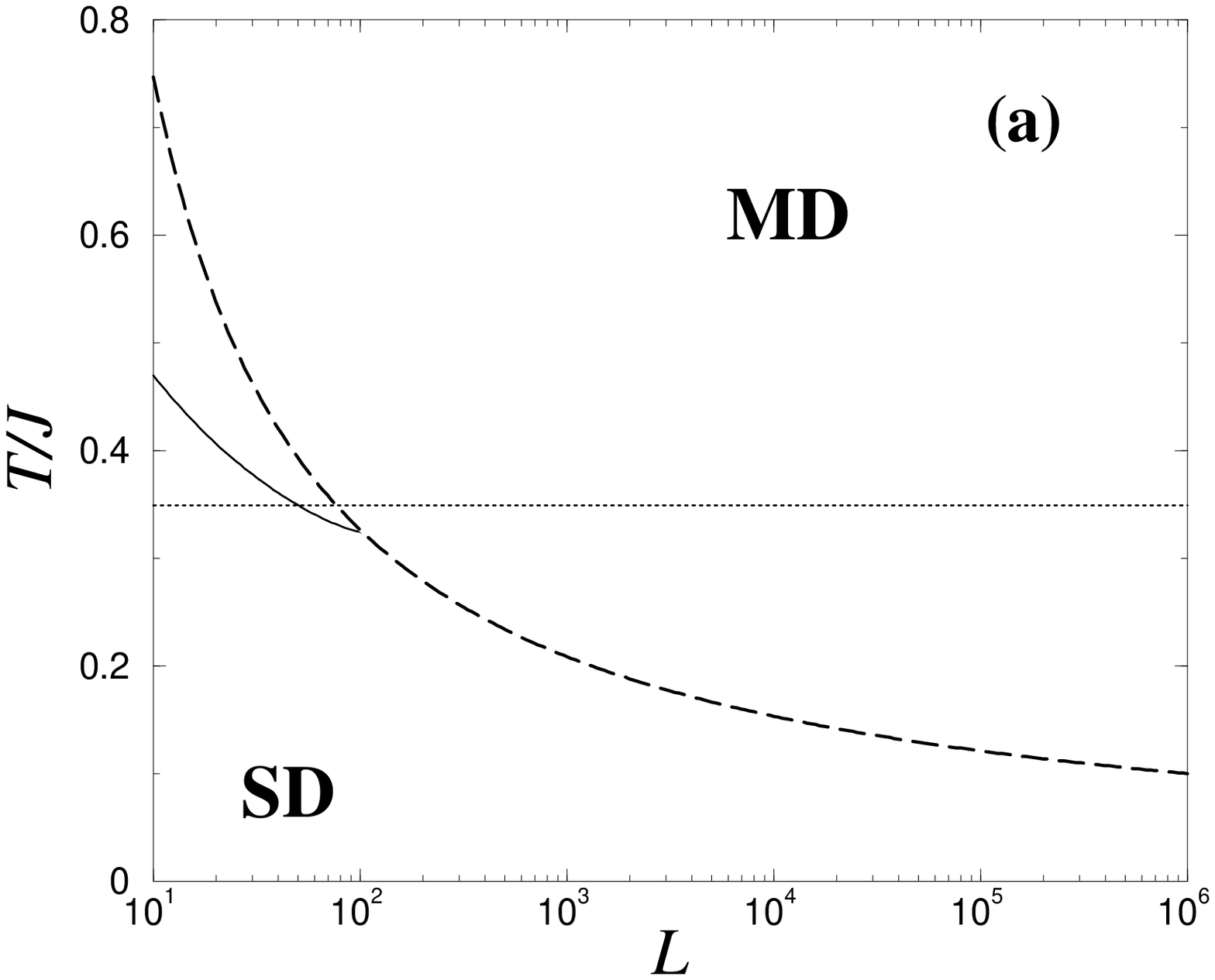}
\includegraphics{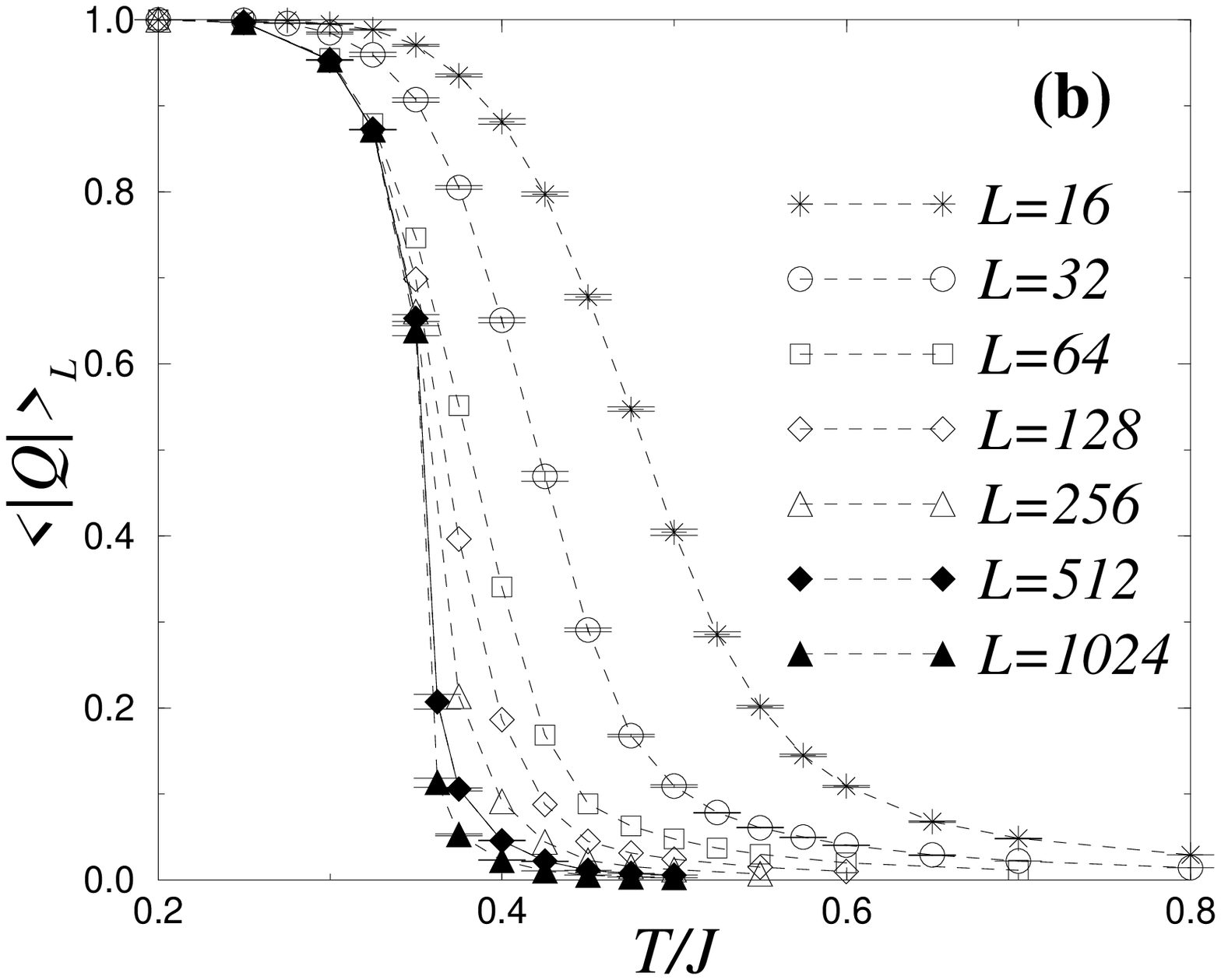}
\includegraphics{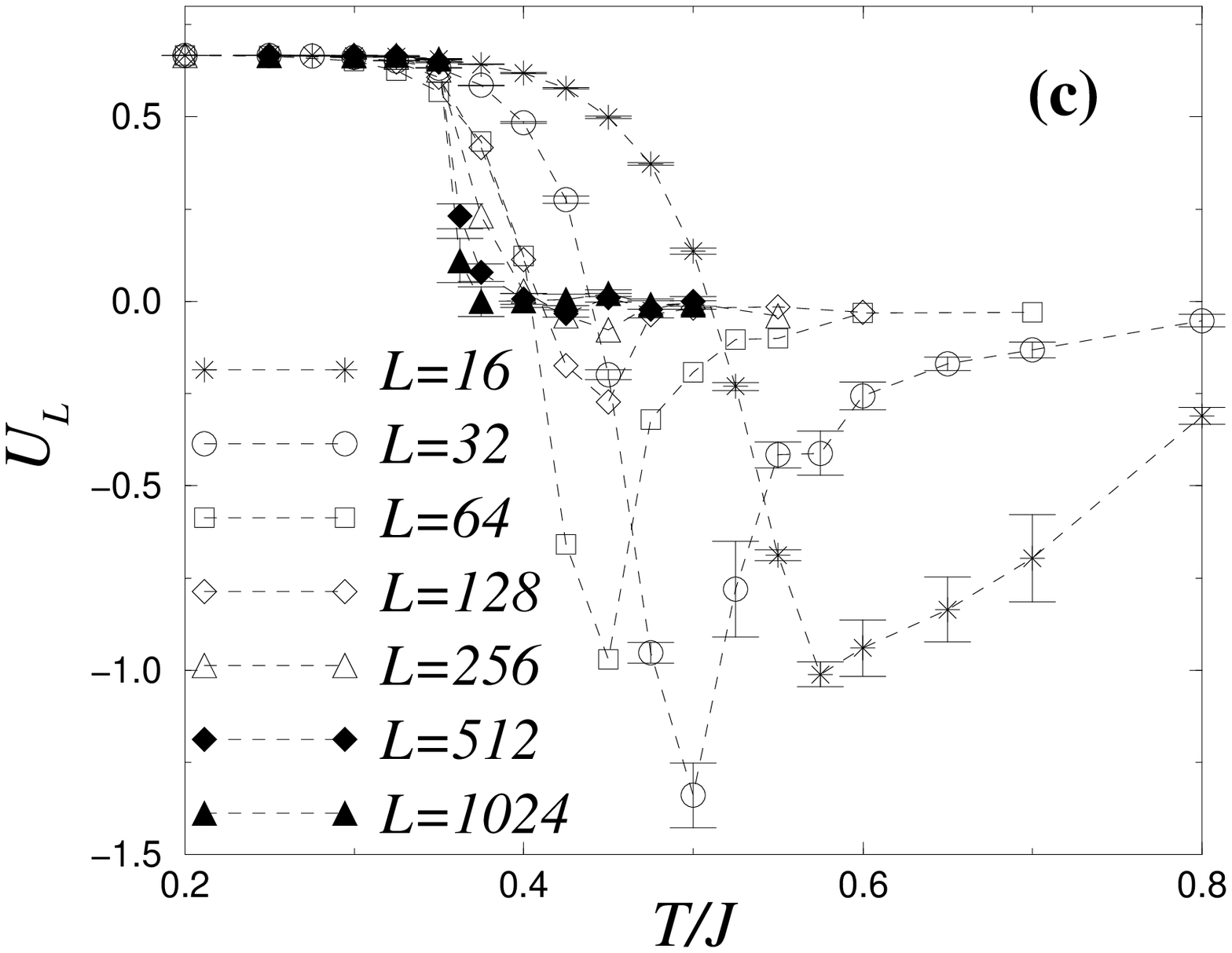}
\vspace*{2.5truecm}
\caption[]{Dependence of the system response on $T$ and $L$ for 
$H$$=$$2.0J$ and $t_{1/2}$$=$50~MCSS. 
(a) 
Metastable dynamic phase diagram analogous to Fig.~\protect\ref{fig1}. 
The different curves have the same interpretations as in that figure. 
The horizontal line corresponds to $T$$=$$0.349J$, where in the MD regime
$\langle\tau(T,H)\rangle \approx t_{1/2} = 50$~MCSS.
(b)
The dynamic order parameter $\langle|Q|\rangle$, shown vs $T$ for $L$ between 
16 and 1024. 
(c)
The fourth-order cumulant ratio $U_L$, shown vs $T$ for $L$ between 
16 and 1024. Note that the dip to negative values disappears as $L$ 
increases beyond 128.
}
\label{fig4}
\end{figure}

\begin{figure}
\vspace*{2.5truecm}
\includegraphics{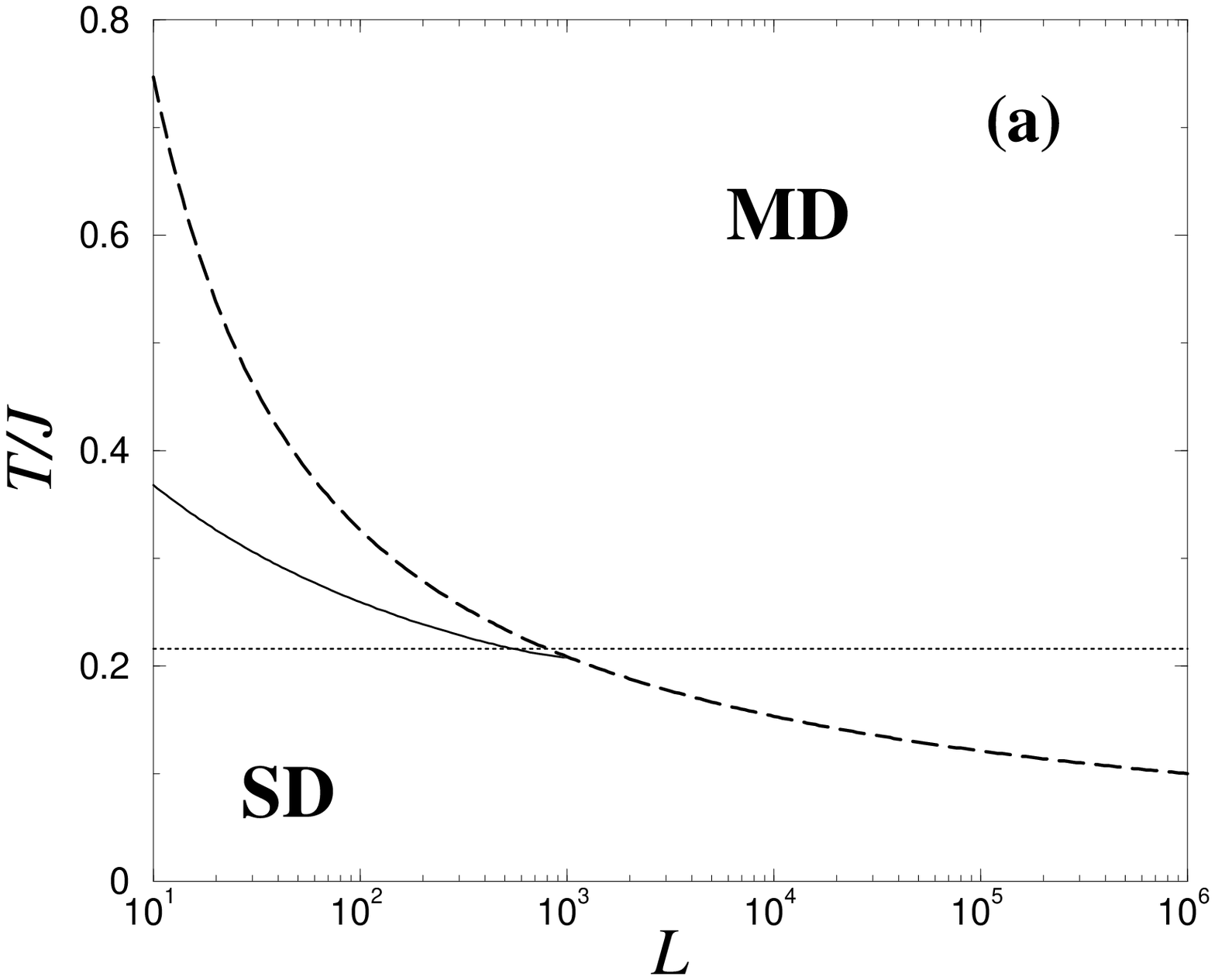}
\includegraphics{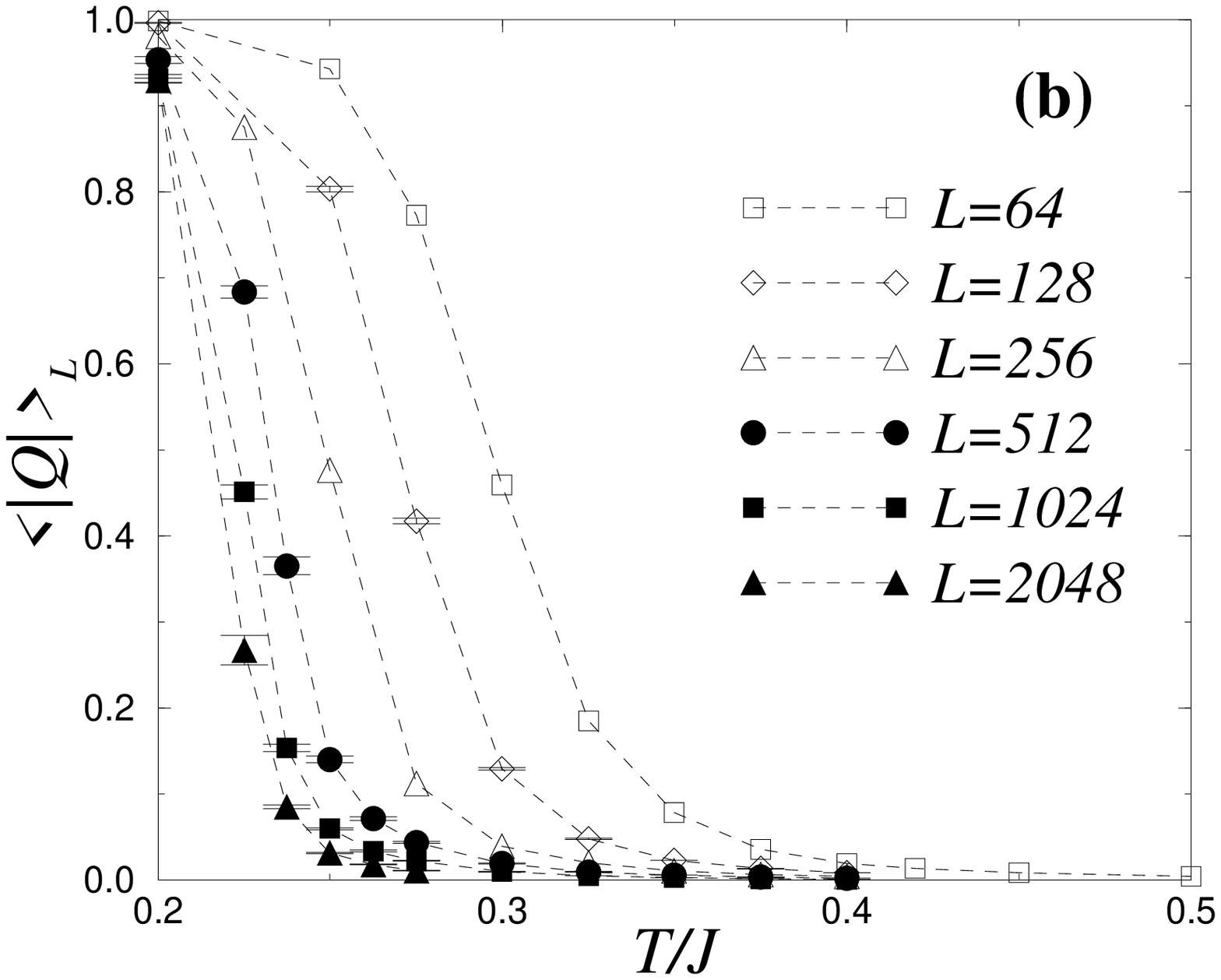}
\includegraphics{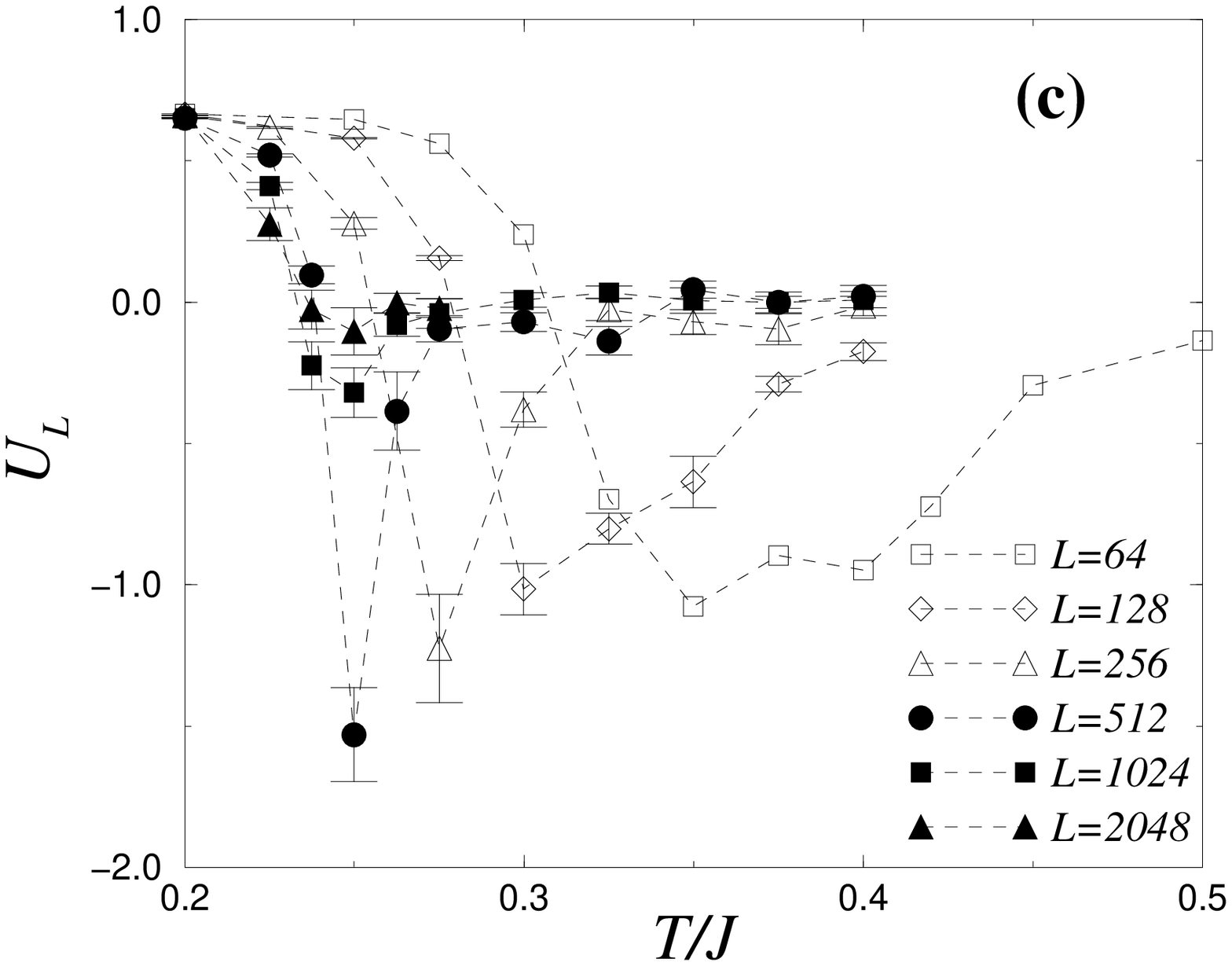}
\vspace*{2.5truecm}
\caption[]{Dependence of the system response on $T$ and $L$ for 
$H$$=$$2.0J$ and $t_{1/2}$$=$500~MCSS. 
(a) 
Metastable dynamic phase diagram analogous to Fig.~\protect\ref{fig1}. 
The different curves have the same interpretations as in that figure. 
The horizontal line corresponds to $T$$=$$0.216J$, where in the MD regime
$\langle\tau(T,H)\rangle \approx t_{1/2} = 500$~MCSS.
(b)
The dynamic order parameter $\langle|Q|\rangle$, shown vs $T$ for $L$ between 
64 and 2048. 
(c)
The fourth-order cumulant ratio $U_L$, shown vs $T$ for $L$ between 
64 and 2048. Note that the dip to negative values gradually disappears 
with increasing $L$. 
}
\label{fig5}
\end{figure}

\begin{figure}
\vspace*{2.5truecm}
\includegraphics{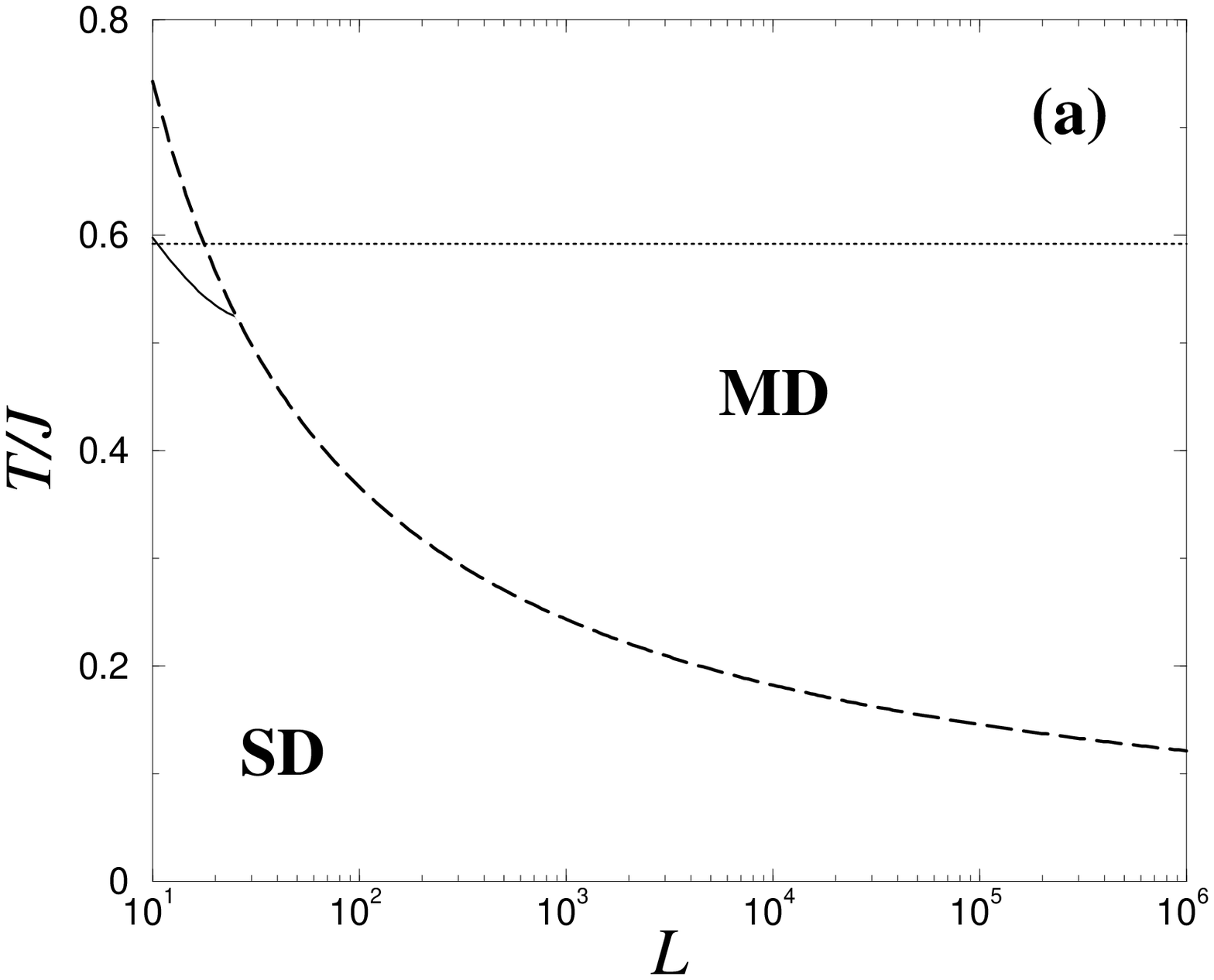}
\includegraphics{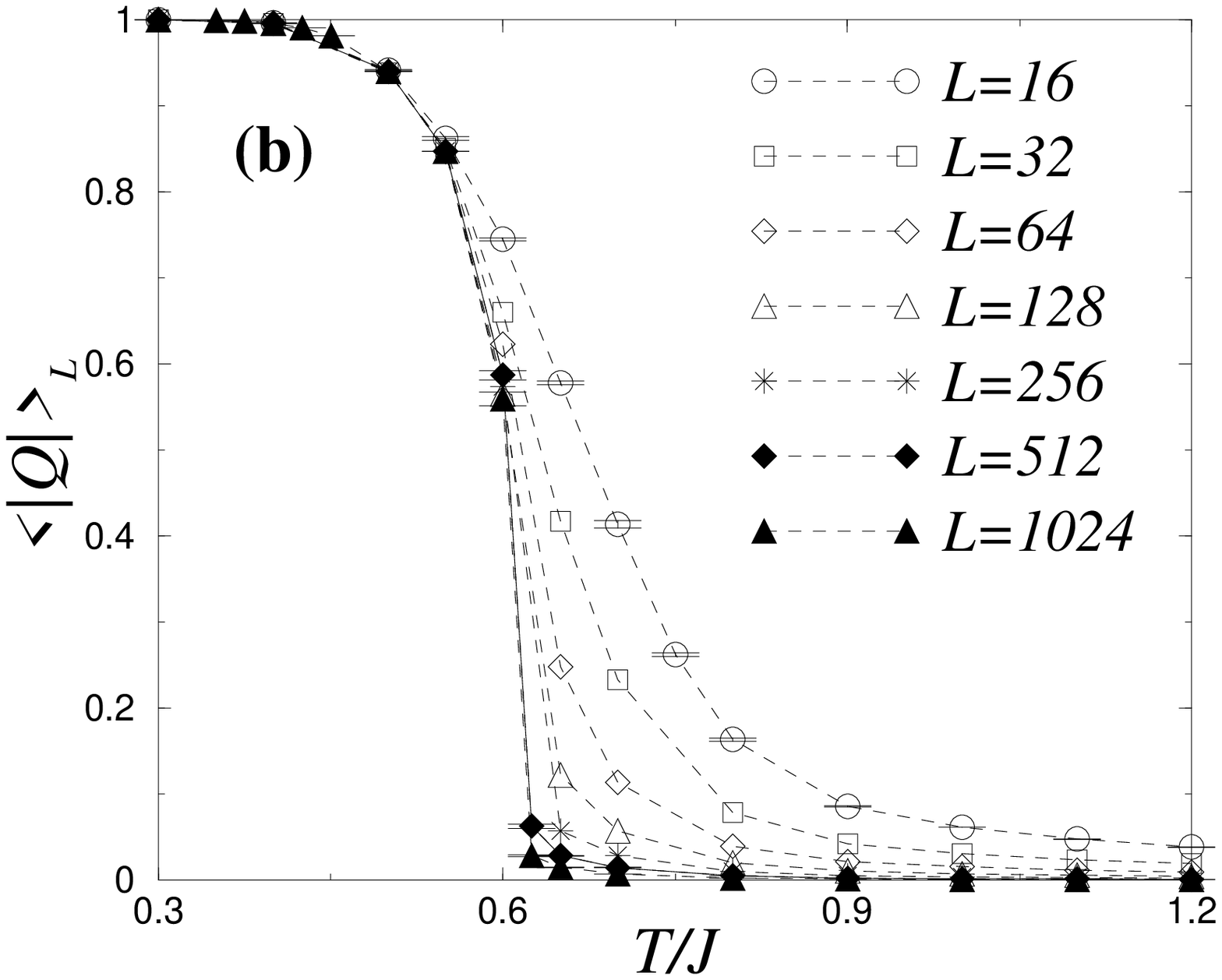}
\includegraphics{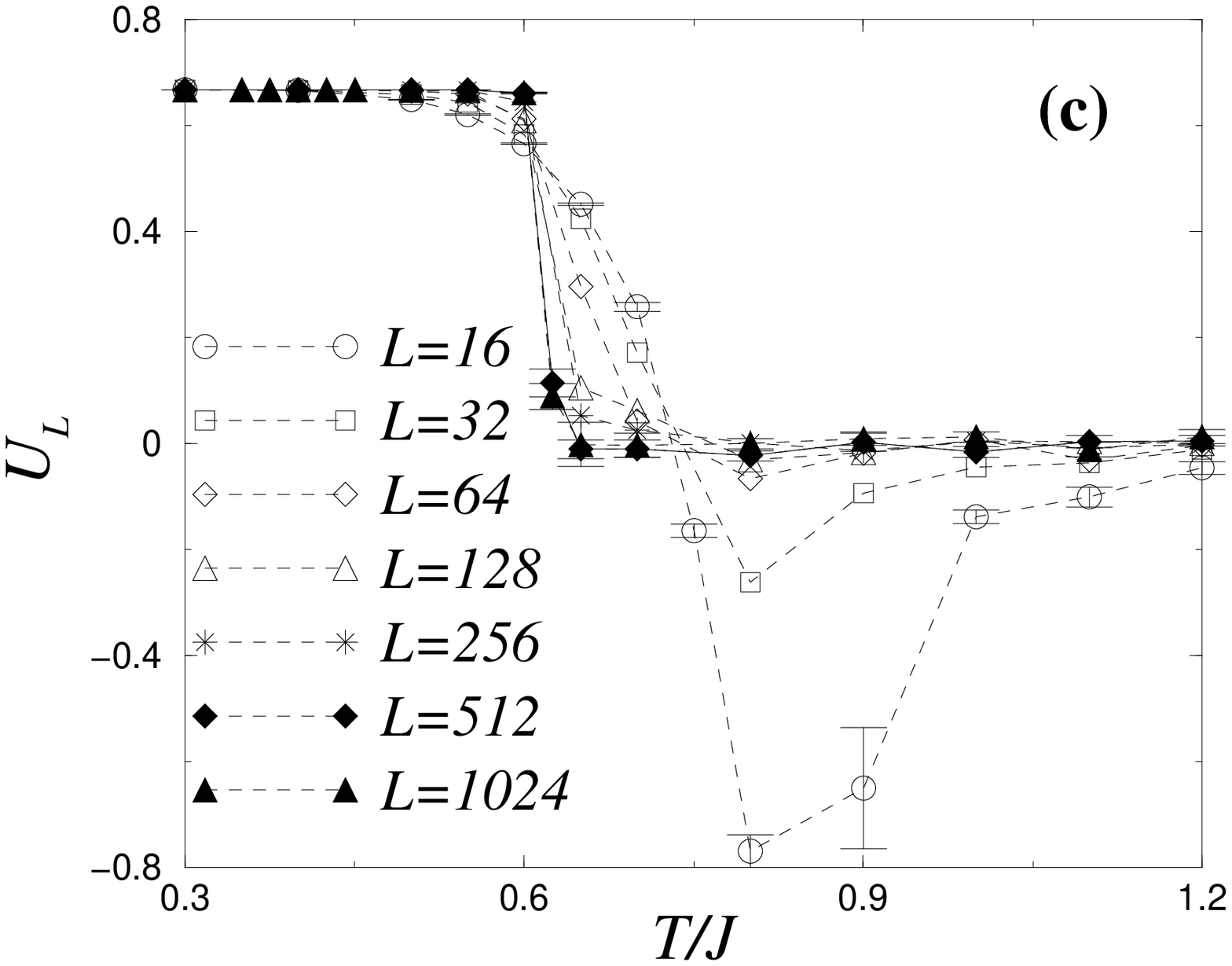}
\vspace*{2.5truecm}
\caption[]{Dependence of the system response on $T$ and $L$ for 
$H$$=$$1.8J$ and $t_{1/2}$$=$20~MCSS. 
(a) 
Metastable dynamic phase diagram analogous to Fig.~\protect\ref{fig1}. 
The different curves have the same interpretations as in that figure. 
The horizontal line corresponds to $T$$=$$0.592J$, where in the MD regime
$\langle\tau(T,H) \rangle \approx t_{1/2} = 20$~MCSS.
(b)
The dynamic order parameter $\langle|Q|\rangle$, shown vs $T$ for $L$ between 
16 and 1024. 
(c)
The fourth-order cumulant ratio $U_L$, shown vs $T$ for $L$ between 
16 and 1024. Note that the dip to negative values disappears as $L$ 
increases beyond 64. 
}
\label{fig6}
\end{figure}

\begin{figure}
\vspace*{2.5truecm}
\includegraphics{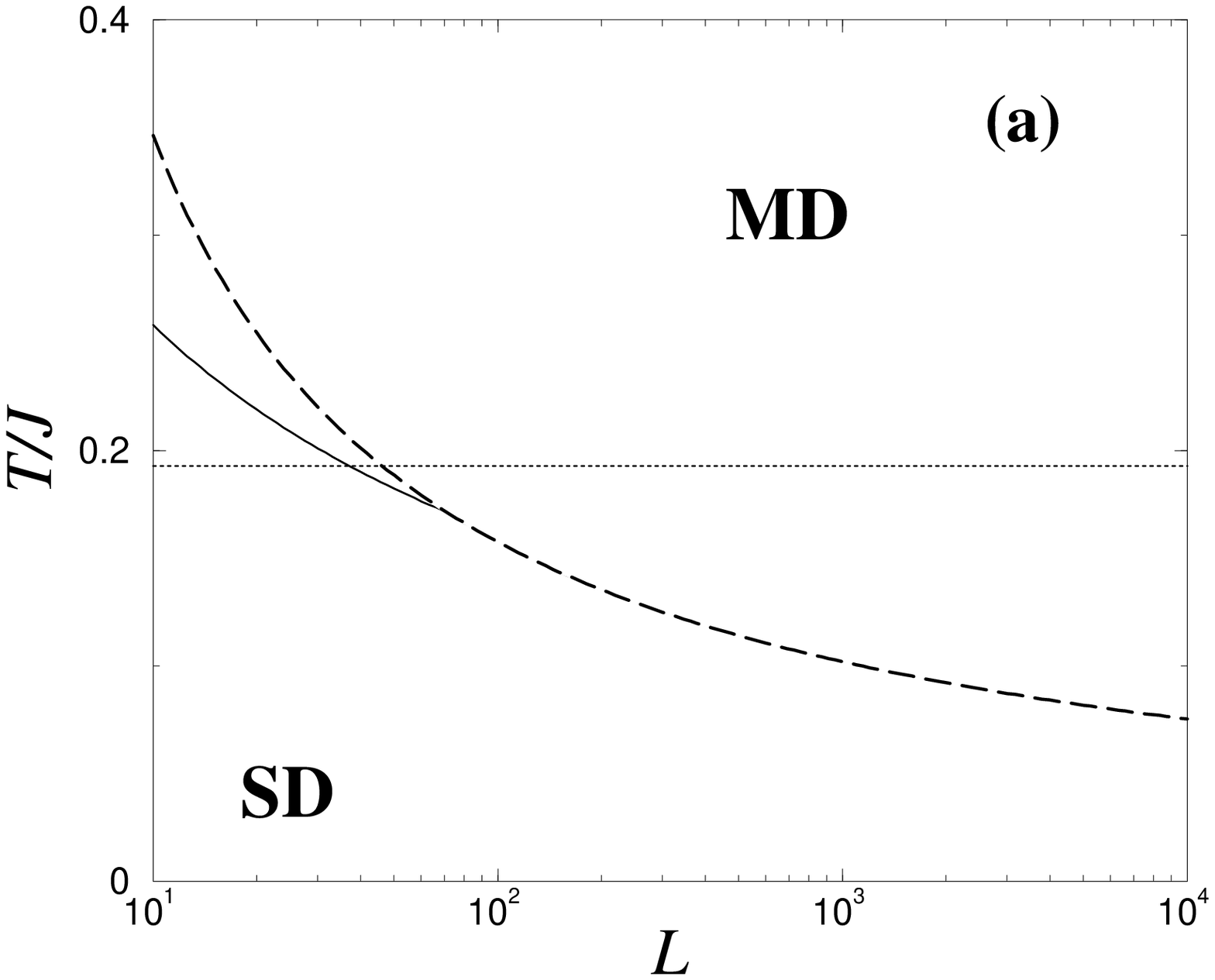}
\includegraphics{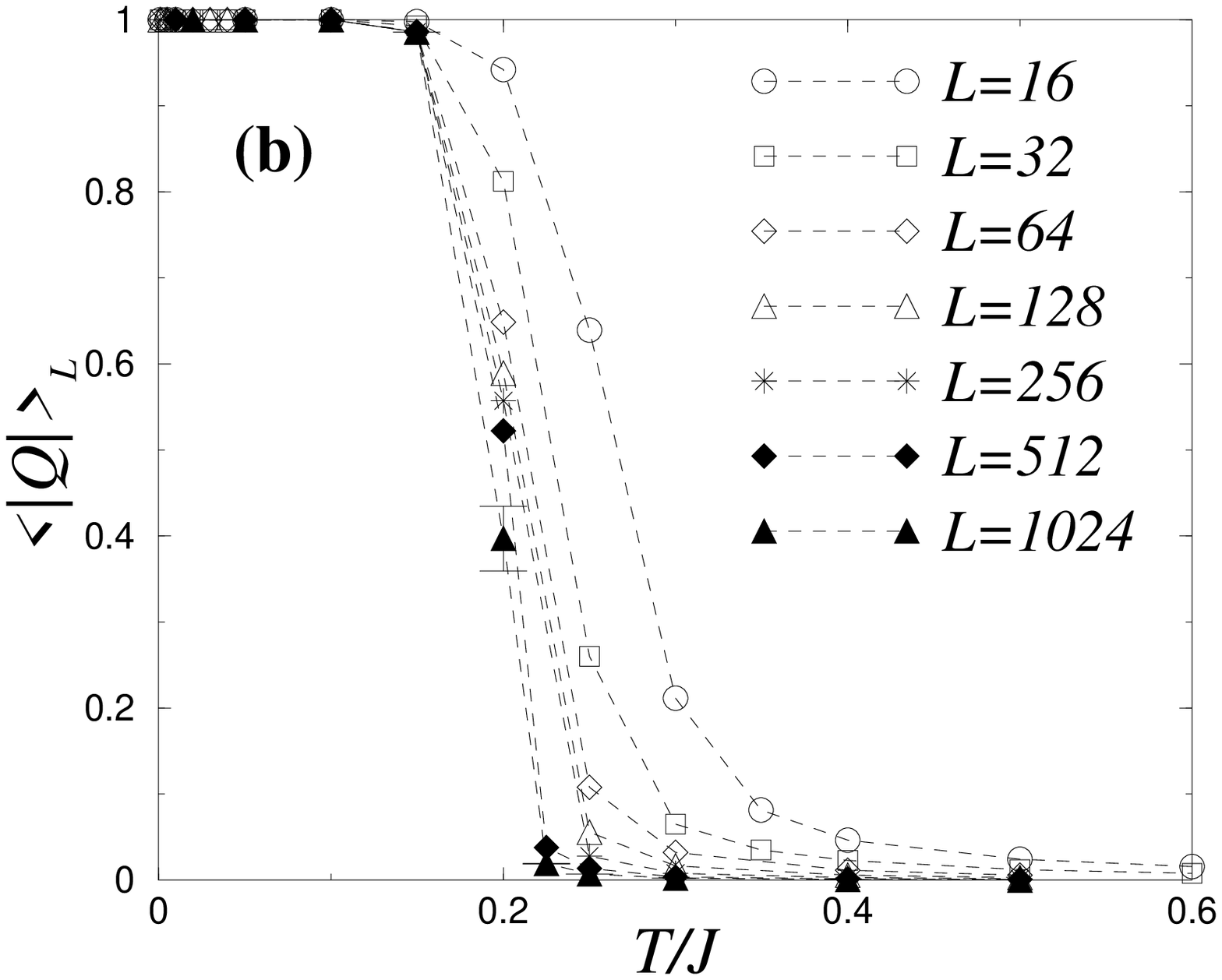}
\includegraphics{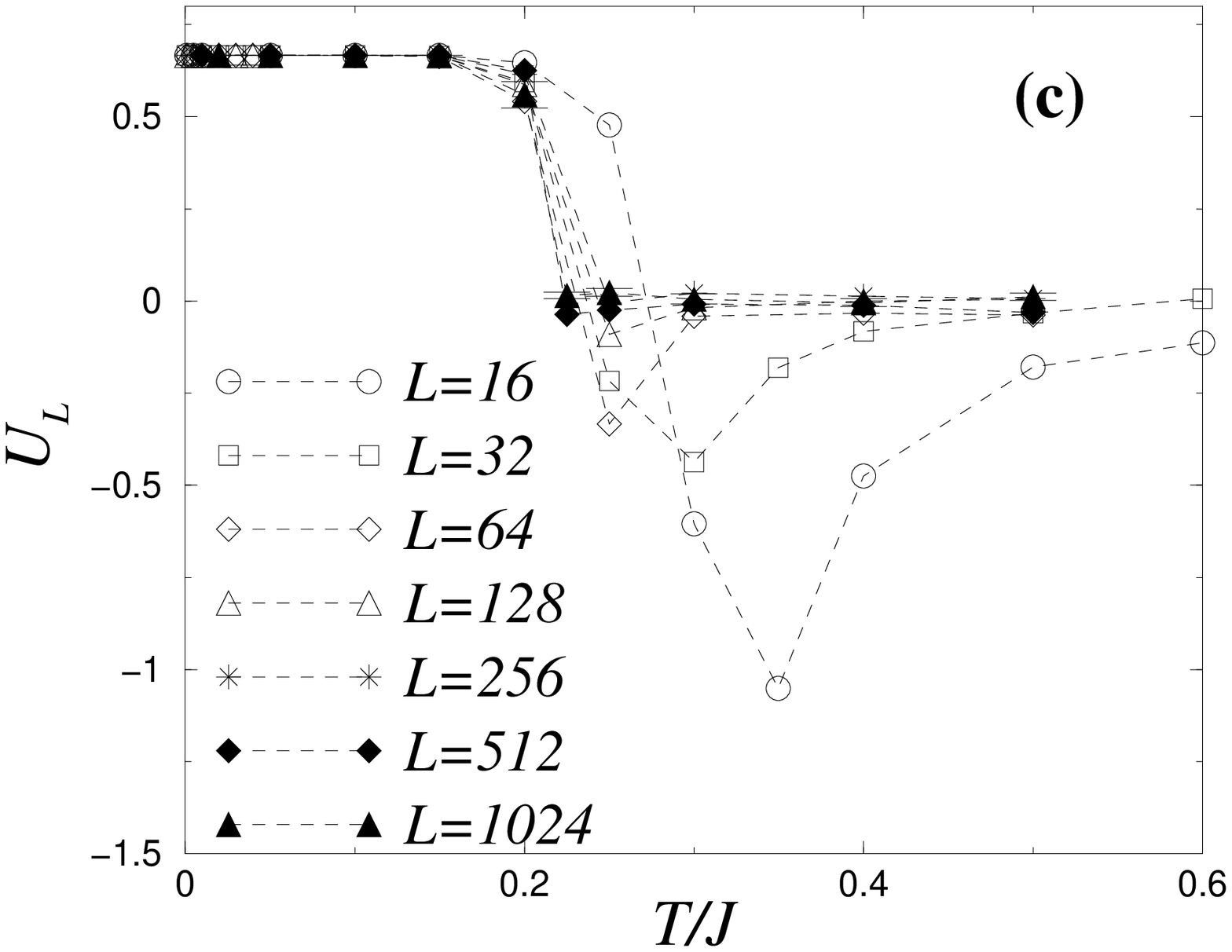}
\vspace*{2.5truecm}
\caption[]{Dependence of the system response on $T$ and $L$ for 
$H$$=$$3.0J$ and $t_{1/2}$$=$20~MCSS. 
(a) 
Metastable dynamic phase diagram analogous to Fig.~\protect\ref{fig1}. 
The different curves have the same interpretations as in that figure. 
The horizontal line corresponds to $T$$=$$0.193J$, where in the MD regime
$\langle\tau(T,H)\rangle \approx t_{1/2} = 20$~MCSS.
(b)
The dynamic order parameter $\langle|Q|\rangle$, shown vs $T$ for $L$ between 
16 and 1024. 
(c)
The fourth-order cumulant ratio $U_L$, shown vs $T$ for $L$ between 
16 and 1024. Note that the dip to negative values disappears with increasing 
$L$.  
}
\label{fig7}
\end{figure}

\begin{figure}[t]
\vspace*{2.0truecm}
\includegraphics{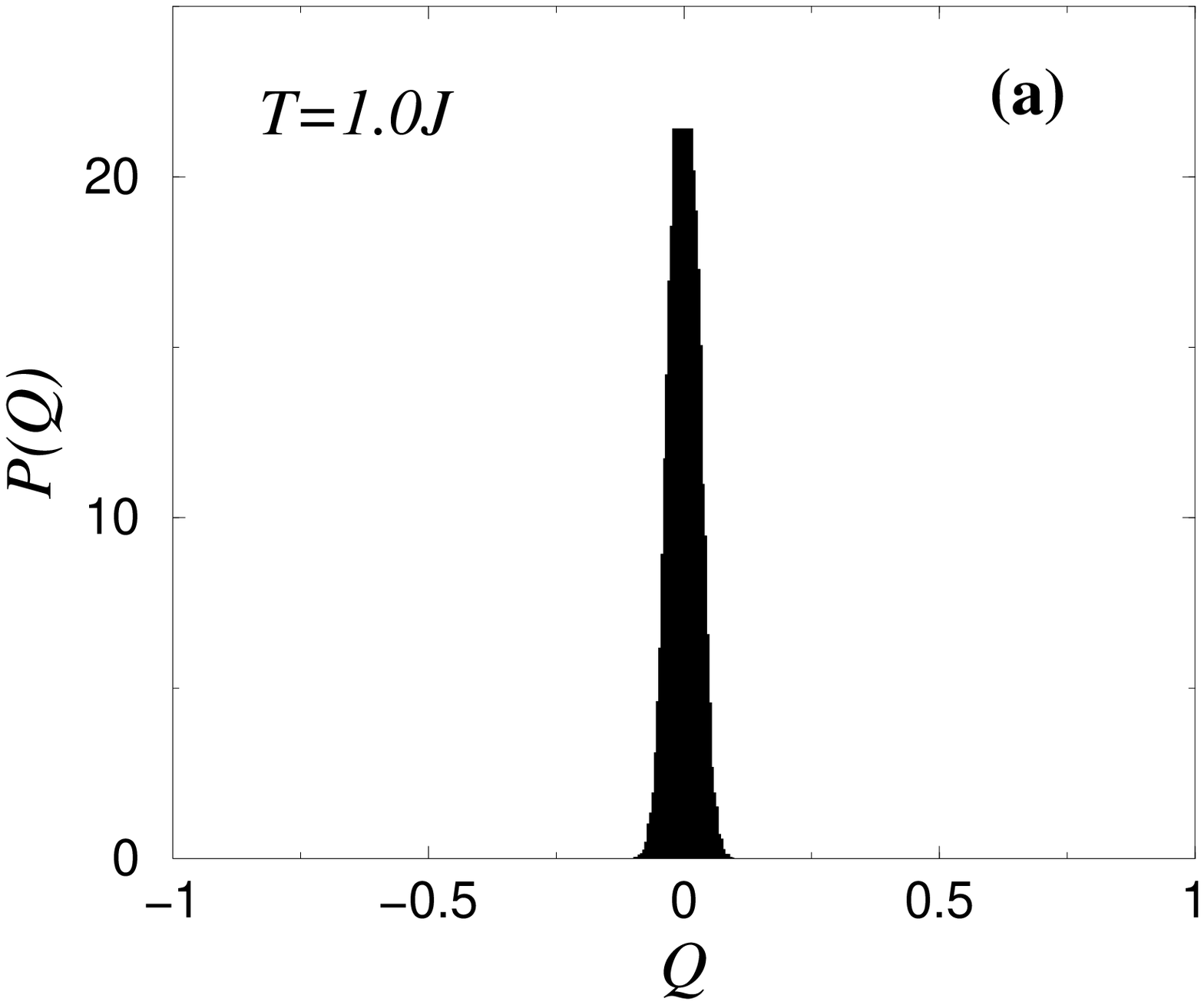}
\includegraphics{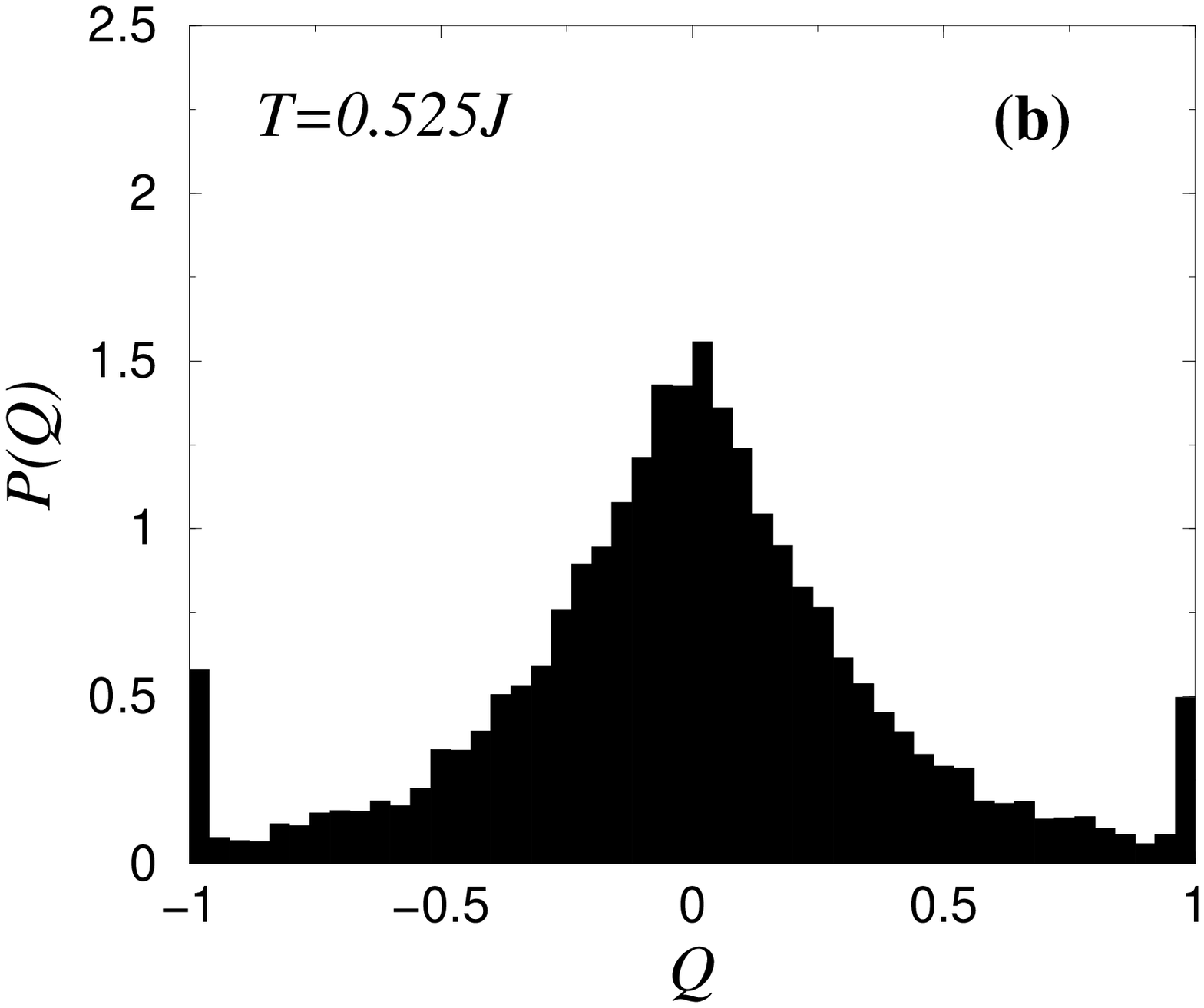}
\includegraphics{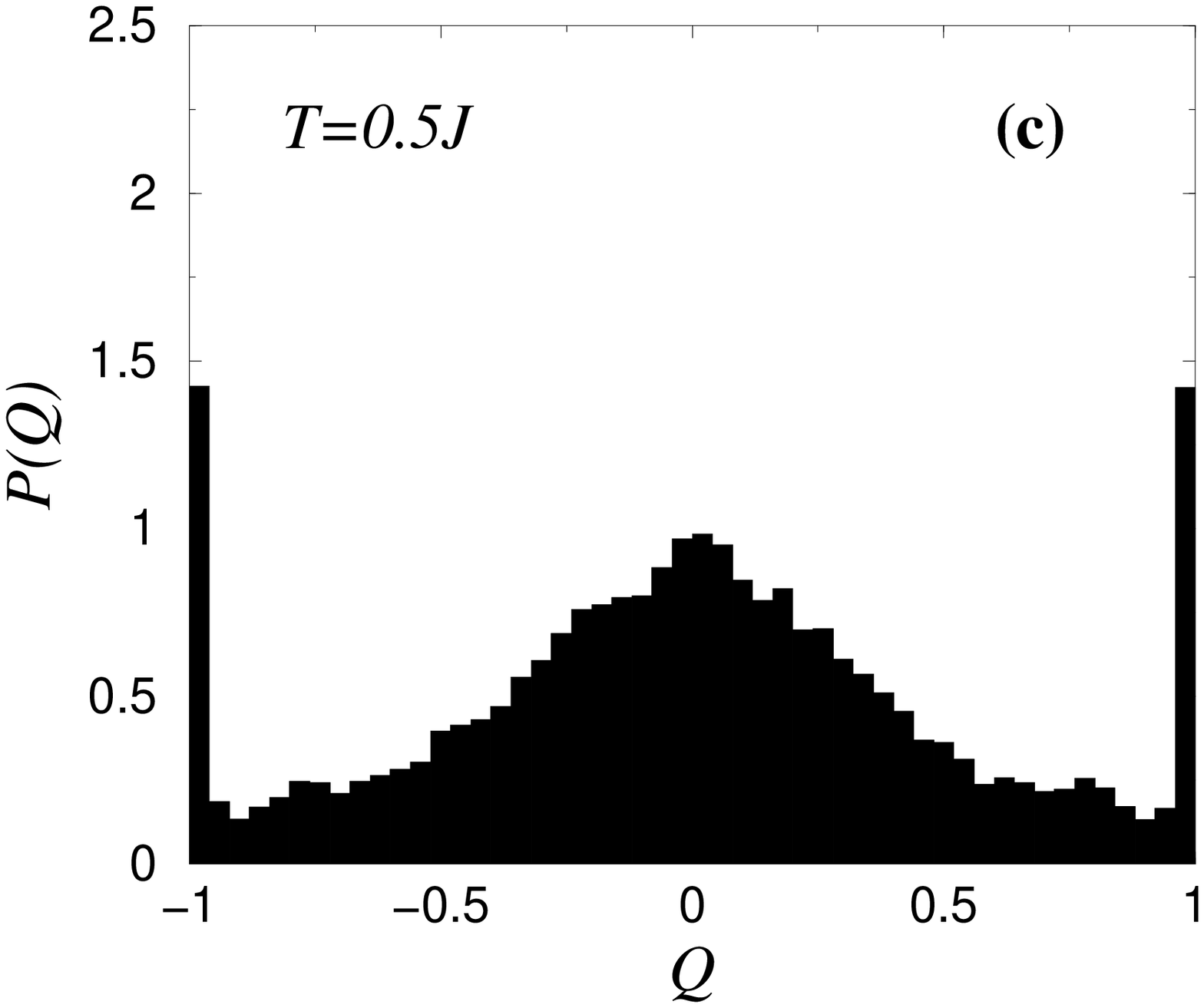}
\includegraphics{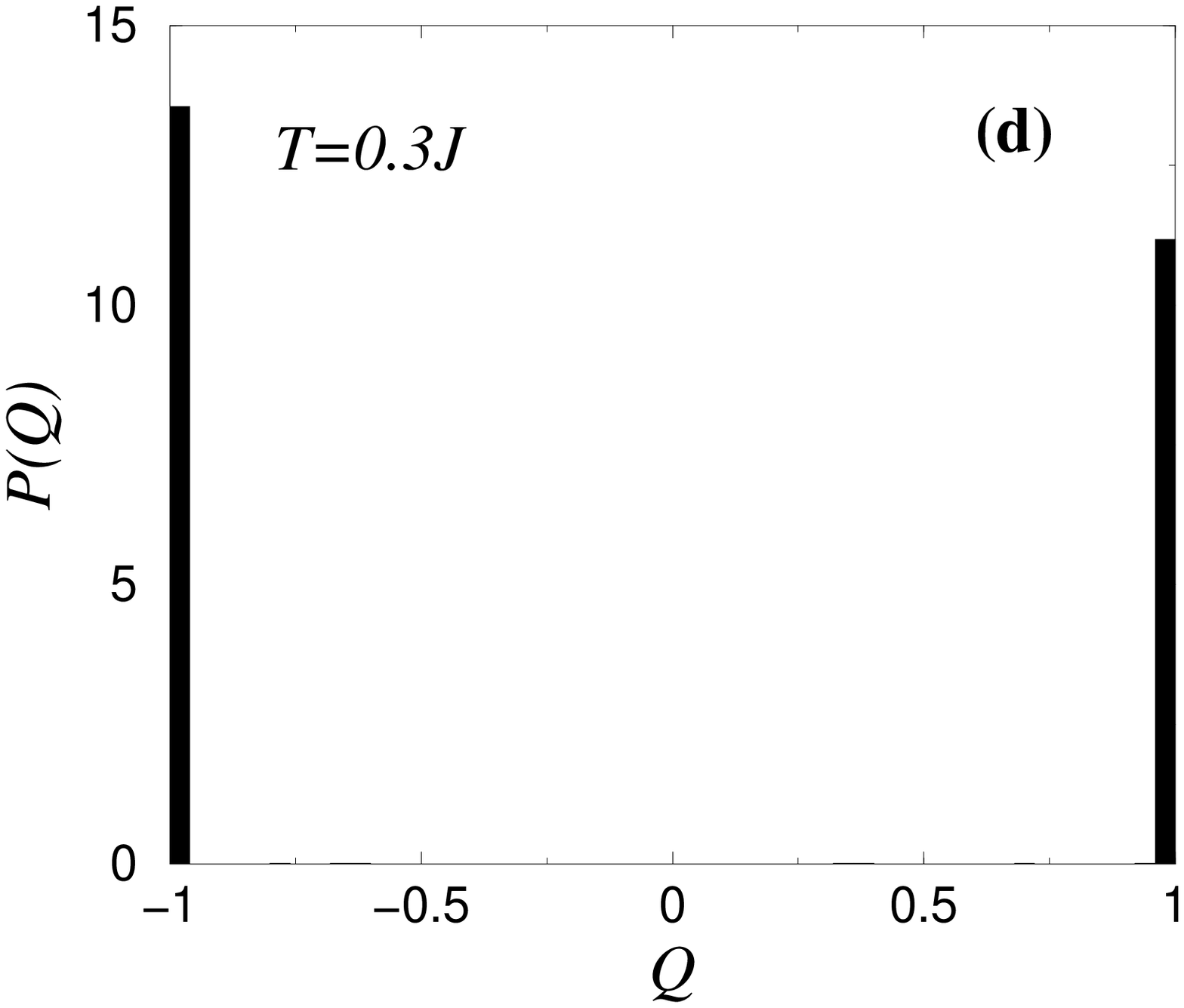}
\vspace*{2.0truecm}
\caption[]{Histograms representing $P(Q)$ in a small system 
($L$=16 with  $H$$=$$2.0J$ and $t_{1/2}$$=$$50$~MCSS) 
for different temperatures.}
\label{fig8}
\end{figure}

Even for relatively small systems, $L$$=$$16$--$128$, at
sufficiently high temperatures the underlying metastable decay mode is MD, as
illustrated in the phase diagram in Fig.~\ref{fig4}(a). Then the
system magnetization $m(t)$ follows a symmetric limit cycle 
[see Fig.~\ref{fig2}(a)]. Consequently, the order-parameter distribution $P(Q)$
is sharply peaked about zero [Fig~\ref{fig8}(a)], yielding
$\langle|Q|\rangle_L$$\sim$$0$ up to finite-size effects, as
shown in Fig.~\ref{fig4}(b). Correspondingly, the fourth-order
cumulant $U_L$ is close to $0$, as shown in Fig~\ref{fig4}(c).

\begin{figure}[t]
\vspace*{2.0truecm}
\includegraphics{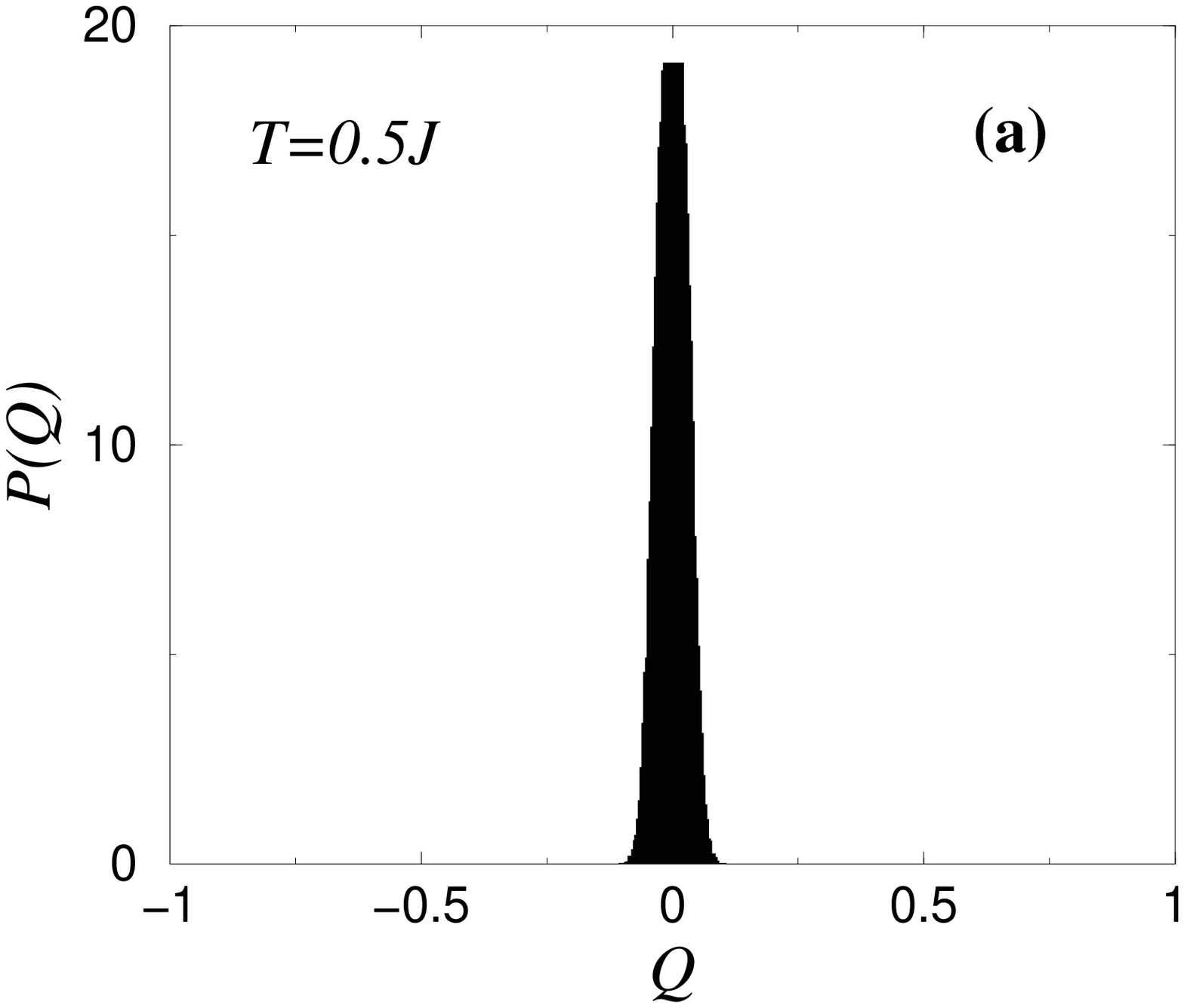}
\includegraphics{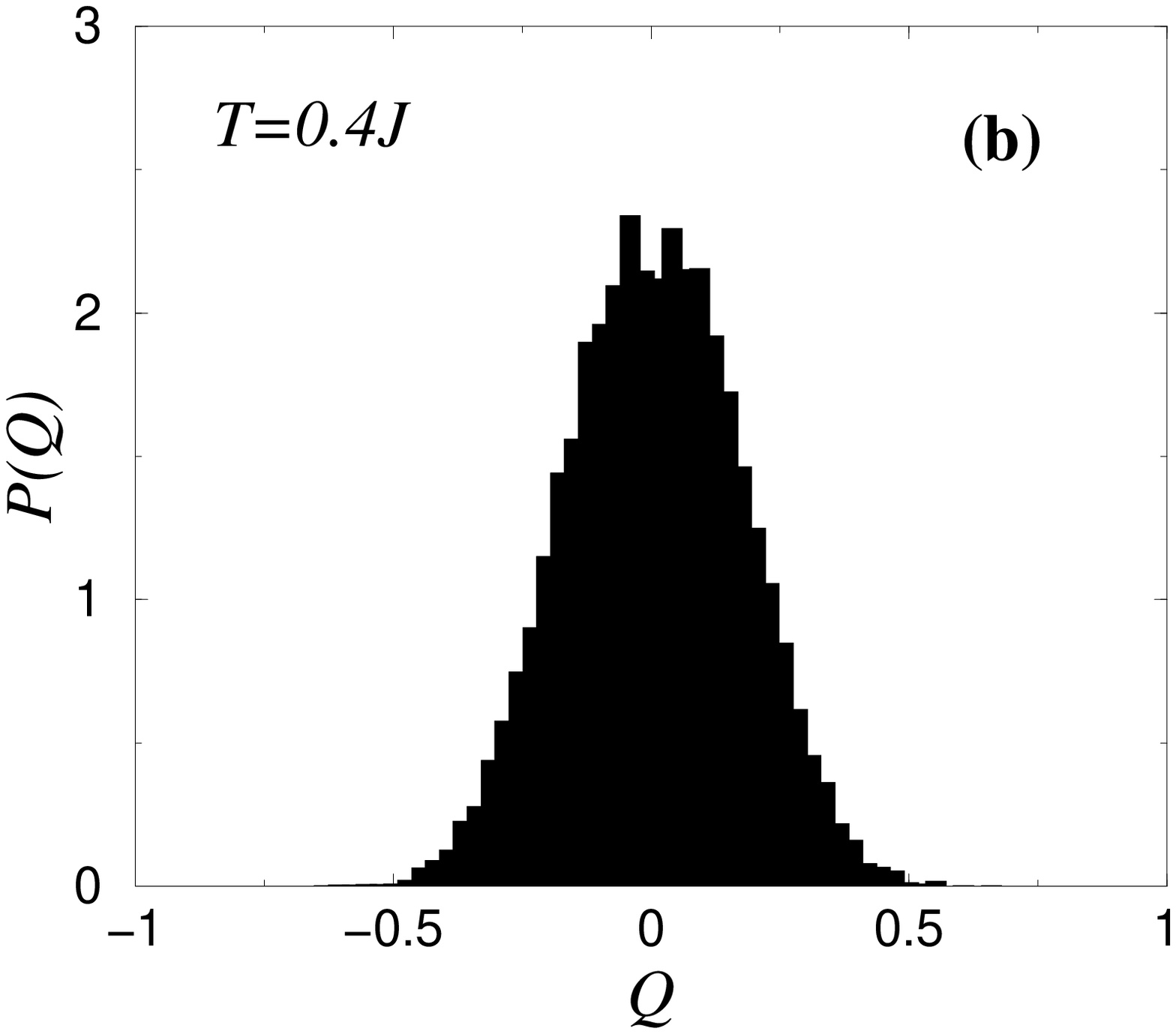}
\includegraphics{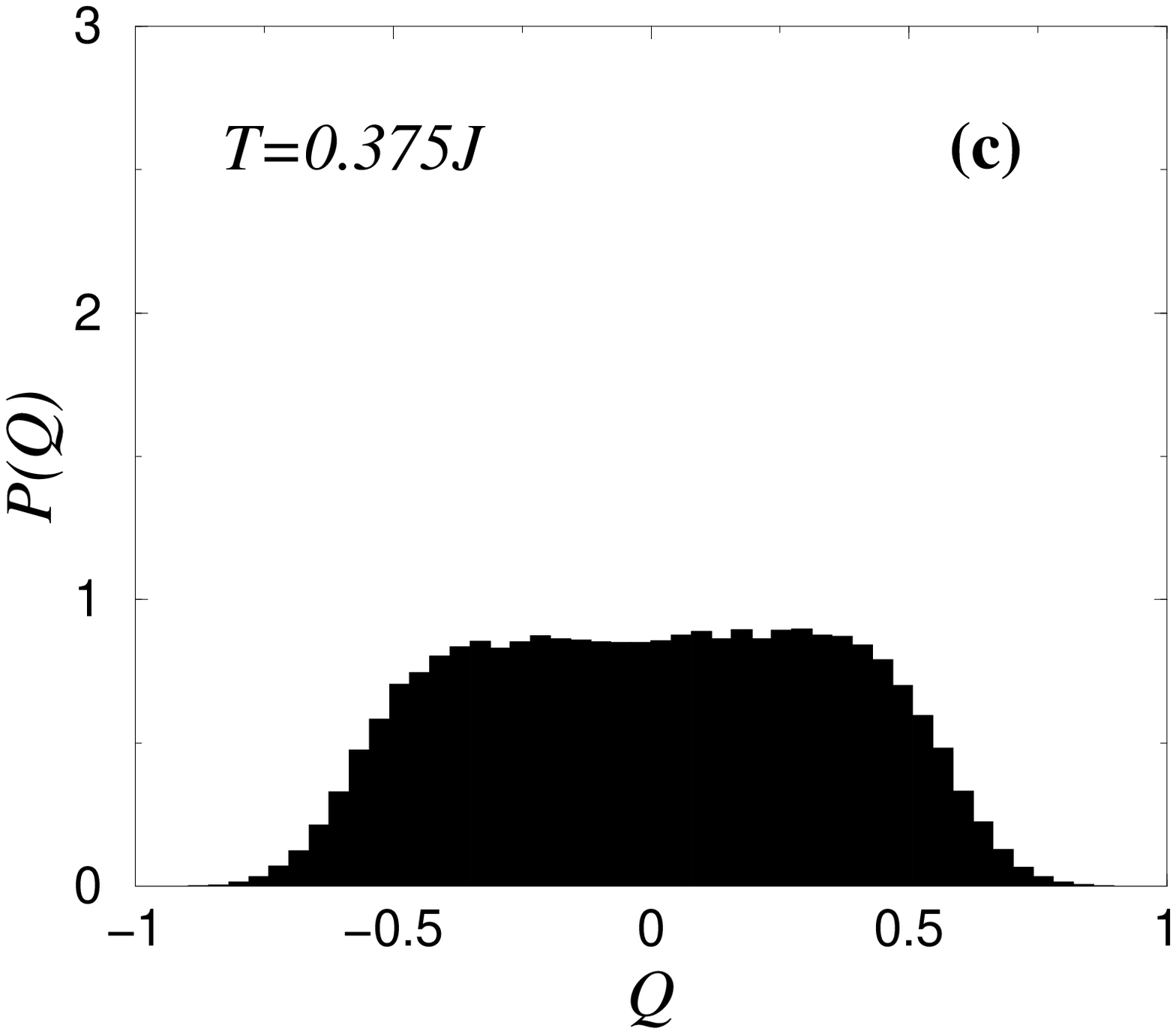}
\includegraphics{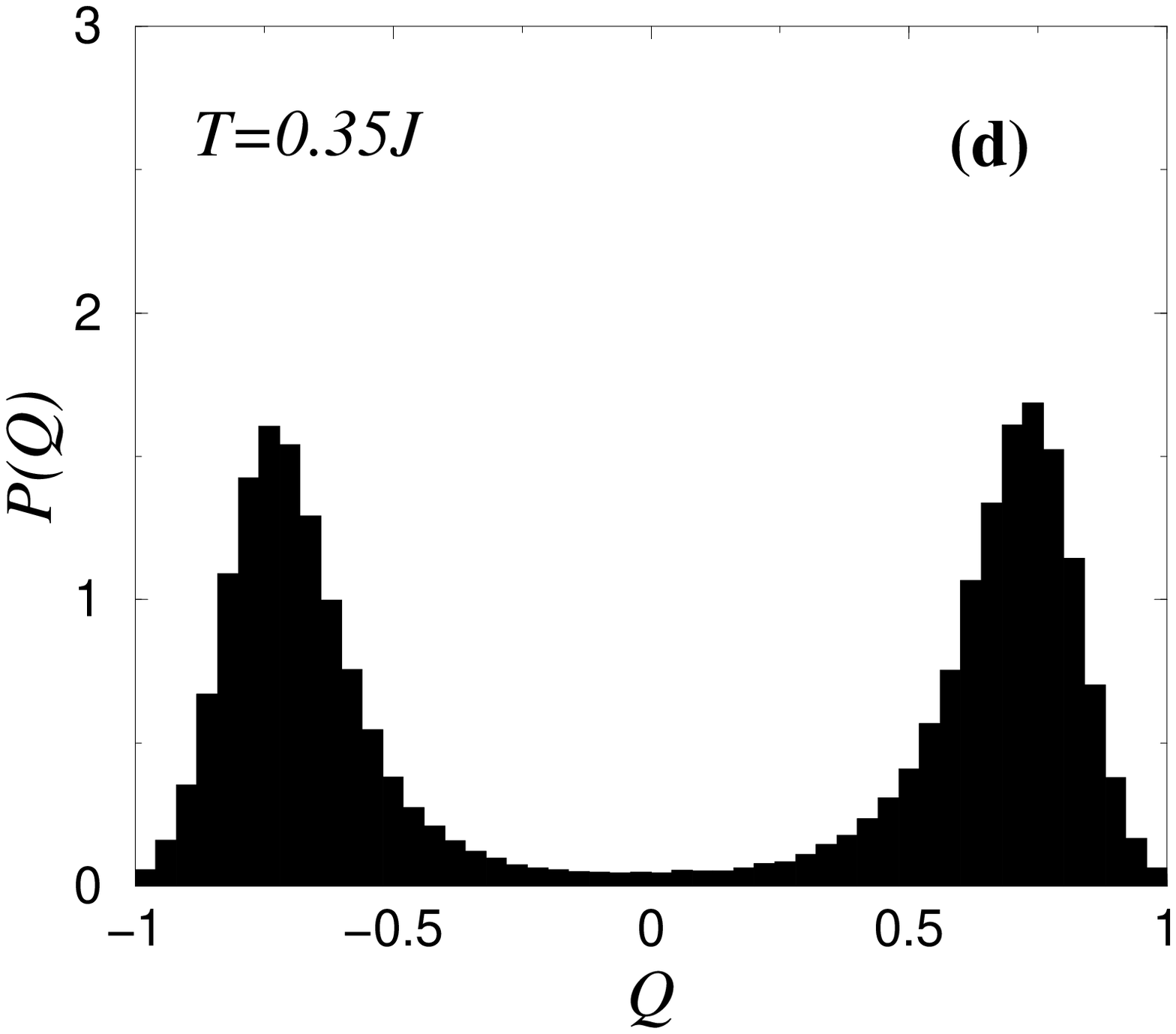}
\vspace*{2.0truecm}
\caption[]{Histograms representing $P(Q)$ in a large system 
($L$=180 with $H$$=$$2.0J$ and $t_{1/2}$$=$$50$~MCSS) 
for different temperatures.}
\label{fig9}
\end{figure}

\begin{figure}[t]
\vspace*{2.5truecm}
\includegraphics{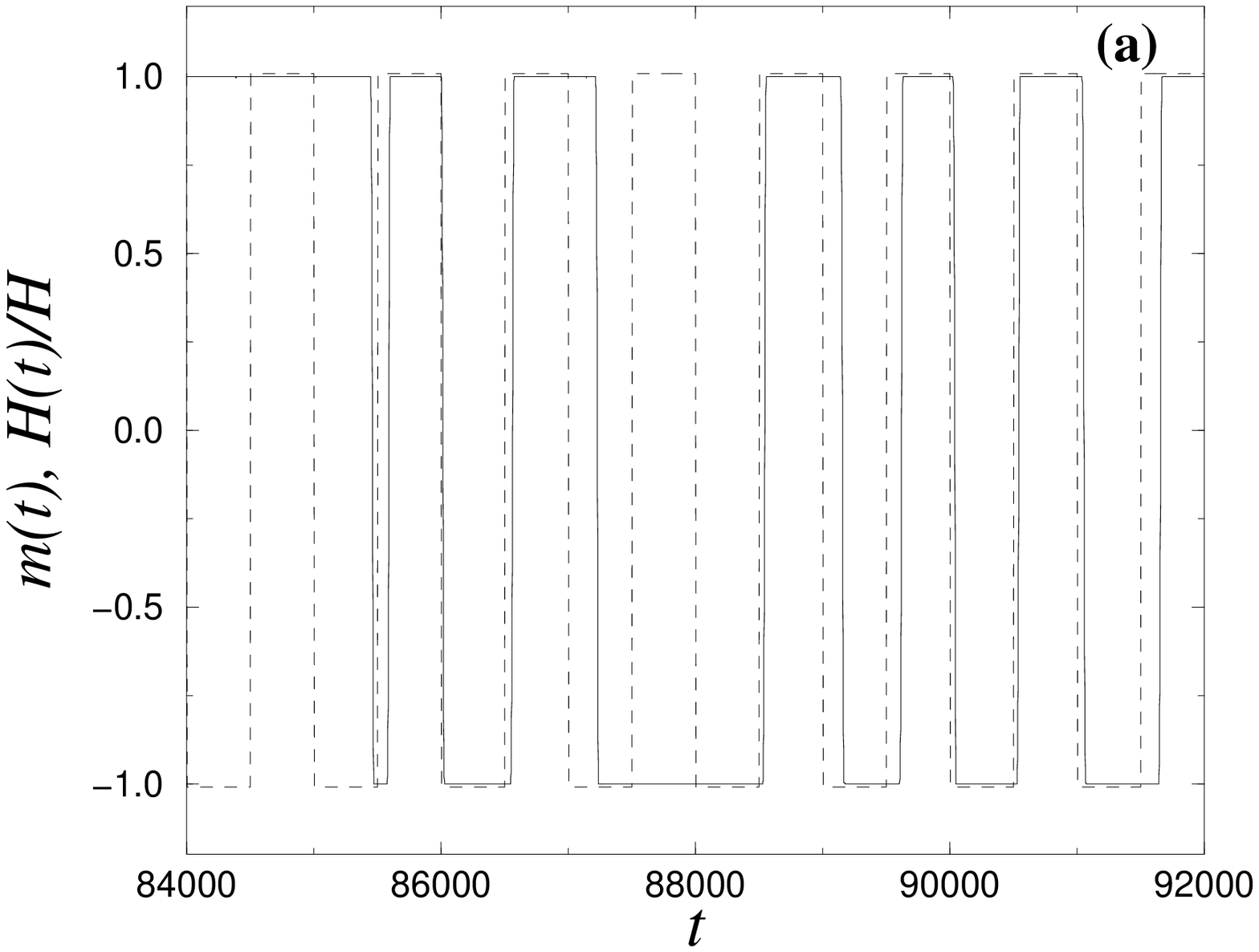}
\includegraphics{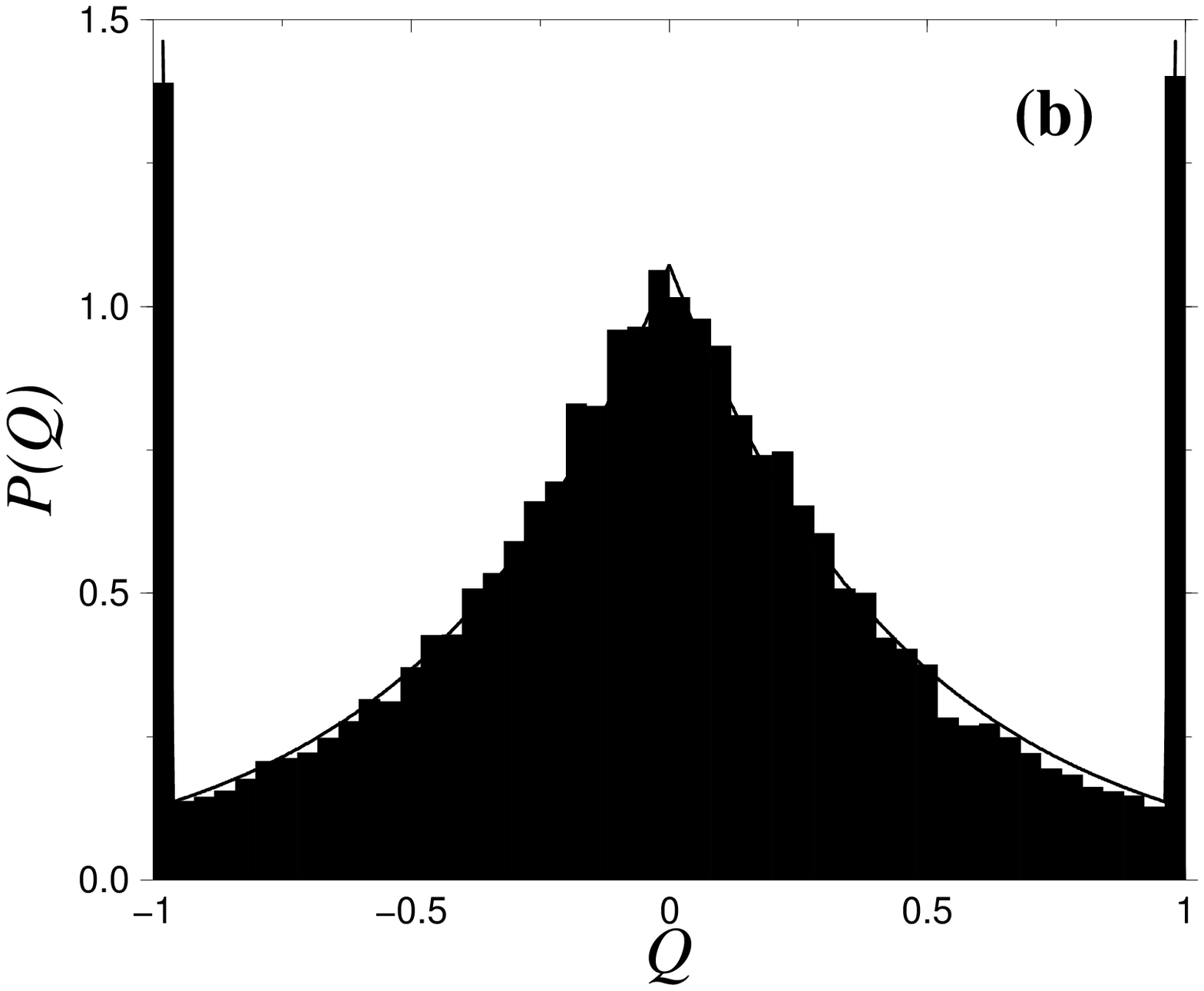}
\vspace*{2.5truecm}
\caption[]{
Order-parameter statistics for 
$H$$=$$2.0J$, $T$$=$$0.34J$, and $L$$=$$32$, which yield 
$\langle\tau(T,H)\rangle_L$$=$$233$~MCSS, 
with $t_{1/2}$$=$$500$~MCSS. 
These are the same parameters as for Fig.~\protect\ref{fig11}. 
For these parameters the system is in the 
stochastic regime, and $t_{\rm g}$ is negligible compared to $t_{1/2}$ 
and $\langle\tau(T,H)\rangle_L$.
(a) 
A short segment of the magnetization time series. 
The interpretation of the line types is the same as in Fig.~\ref{fig2}. 
(b) 
Comparison of the simulated (histogram) 
and the analytic [solid curve, Eq.~(\protect\ref{P_Q1})] 
order-parameter distributions.}
\label{fig10}
\end{figure}

As the temperature is lowered beneath $T_{\rm DSP}(L)$, 
the underlying decay mode crosses over to the SD regime, 
and the magnetization switching becomes stochastic as shown in 
Fig.~\ref{fig2}(b). The order-parameter distribution then has three peaks: two
extremely narrow peaks near $Q$$=$$\pm 1$ and a rather wide one
centered at $Q$$=$$0$ [see Fig.~\ref{fig8}(b) and (c)]. The peaks near $\pm
1$ represent the periods during which the magnetization does not switch, while 
the peak centered on zero represents the periods during which it
switches at least once. The large width of this central
peak is the result of the square-wave shape of the applied field,
which results in an exponential probability density for the
switching process. A sinusoidal field would yield a
distribution more sharply peaked about zero,  
since, in that case, the switching almost always occurs when the external
field assumes its maximum magnitude \cite{ACHA99,SIDES96,SIDES97,SIDES98b}.
The generic feature in the stochastic regime,  regardless of the shape
of the driving field, is the multiple-peak
structure. For the square-wave field used in this paper one can obtain
(see Appendix~\ref{sec:pq}) an analytic approximation for $P(Q)$ in
the regime where 
$t_{1/2}$$\gg$$t_g$ [Fig.~\ref{fig10}(a)]
\begin{equation}
P(Q) =
\frac{e^{-\Theta}}{2}\delta(Q+1) +
\frac{\Theta}{2}e^{-\Theta |Q|} +
\frac{e^{-\Theta}}{2}\delta(Q-1) \;.
\label{P_Q1}
\end{equation}
Equation (\ref{P_Q1}) for $P(Q)$ is compared with simulation data in
Fig.~\ref{fig10}(b) for a system with $L$=32 at 
$H$$=$$2.0J$ and $T$$=$$0.34J$, for which 
$\langle\tau(T,H)\rangle_L$$=$$233$~MCSS. 
(The subscript $L$ in $\langle\tau(T,H)\rangle_L$ is included as a reminder 
that the metastable lifetime depends on $L$ in the SD regime.)
The half-period is $t_{1/2}$$=$$500~{\rm MCSS}~ \gg t_{\rm g}$.
This comparison contains {\em no} fitting
parameters: the average metastable lifetime of the
underlying metastable decay, $\langle\tau(T,H)\rangle_L$, was measured
in single field-reversal simulations, and its value was used to
determine the scaled half-period,
$\Theta$$=$$t_{1/2}/\langle\tau\rangle$. 
Instead of the delta-functions with amplitude $e^{-\Theta}/2$, the
finite value $e^{-\Theta}/(2\Delta Q)$
was used to make the correspondence with the finite
bin-size $\Delta Q$, employed to build the histogram for $Q$. 
In the stochastic regime the cumulant $U_L$ becomes negative,
reaching a minimum at some temperature, as shown for the smaller values 
of $L$ in Fig.~\ref{fig4}(c). 

\begin{figure}
\vspace*{2.0truecm}
\includegraphics{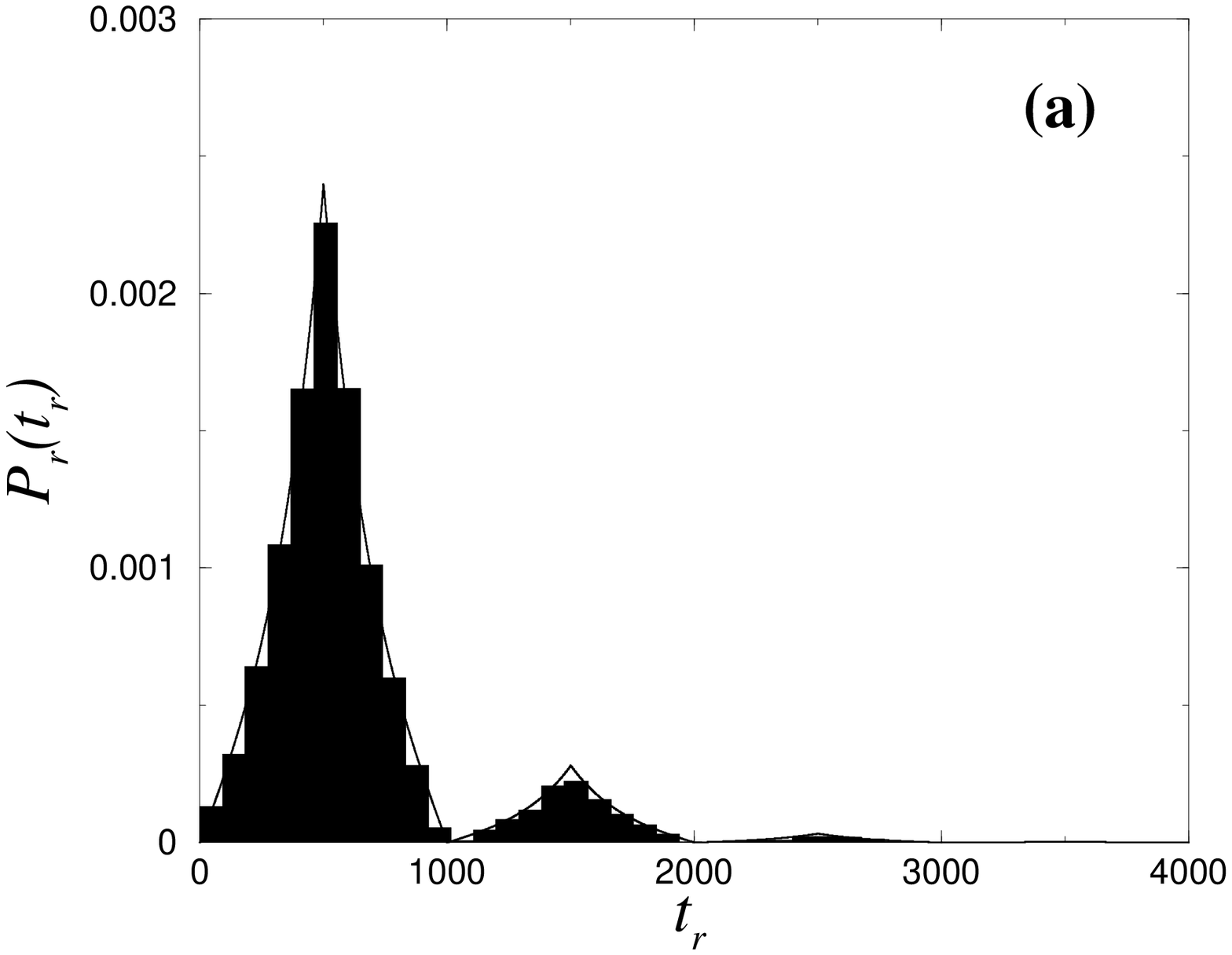}
\includegraphics{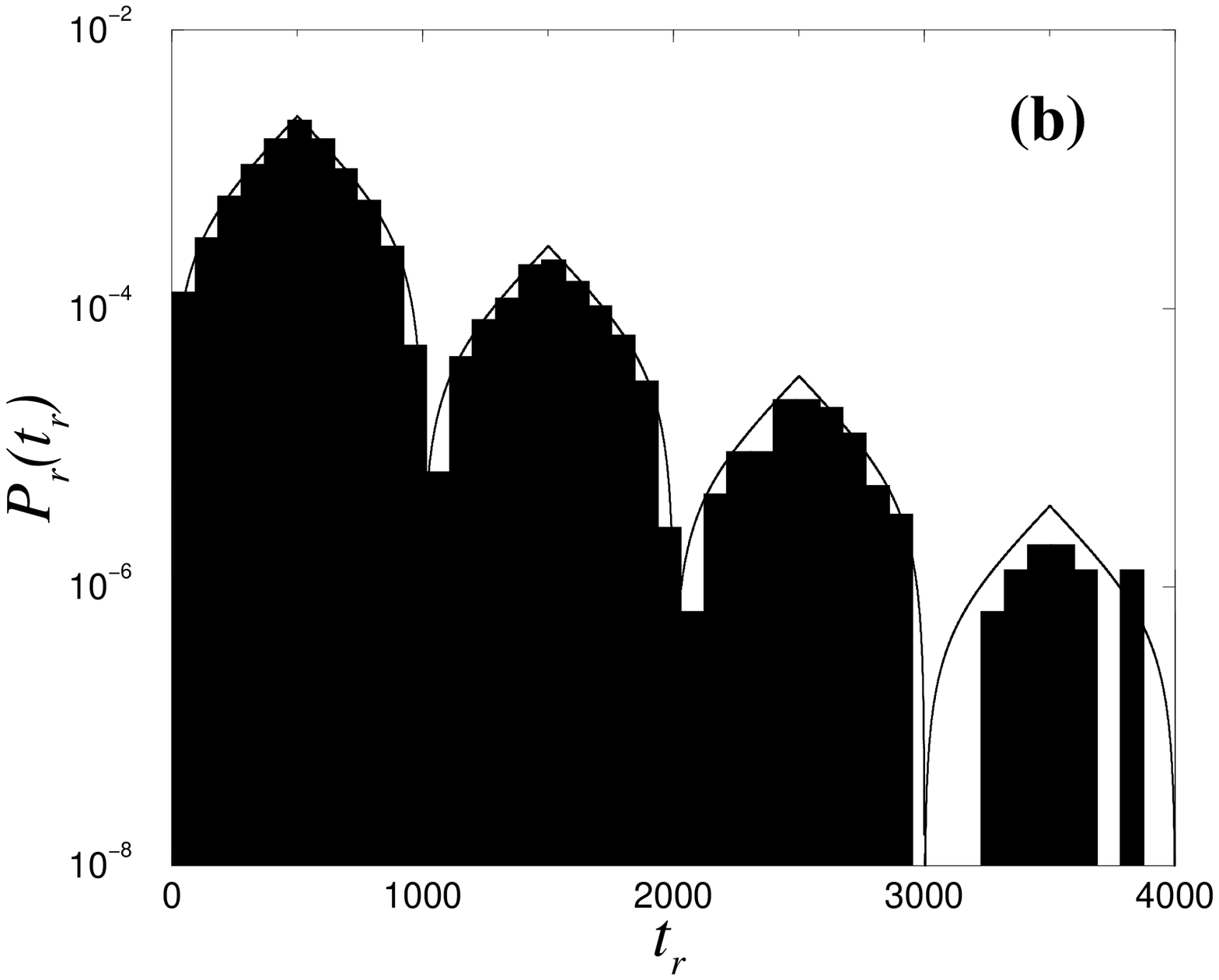}
\vspace*{3.50truecm}
\caption[]{Comparison of the simulated and the analytic 
[Eq.~(\protect\ref{P_res1})]
residence-time distributions, $P_{\rm r}(t_{\rm r})$ in
the stochastic regime where $t_{\rm g} \ll t_{1/2}$. Shown for a small system 
with $L$$=$$32$ at $H$$=$$2.0J$ and $T$$=$$0.34J$, 
where $\langle\tau(T,H)\rangle_L$$=$$233$~MCSS. 
The half-period, $t_{1/2}$$=$500~MCSS, is much longer than the growth time.
(a) Shown on a linear scale. 
(b) The same as (a), using a linear-log scale to emphasize the smaller peaks 
and the exponential dependence of the peak heights on $t_{\rm r}$. 
}
\label{fig11}
\end{figure}

A quantity often used to detect stochastic resonance, is the 
rtd, $P_{\rm r}(t_{\rm r})$ \cite{SIDES97,SIDES98b,SR,JUNG93,GAMM95}, 
the probability 
density for the residence times $t_{\rm r}$ 
between zero-crossings of the magnetization. 
This quantity is shown in Fig.~\ref{fig11} for a system with $L$=32 at 
$H$$=$$2.0J$ and $T$$=$$0.34J$, for which 
$\langle\tau(T,H)\rangle_L$$=$$233$~MCSS.  
The half-period is $t_{1/2}$$=$$500~{\rm MCSS}~ \gg t_{\rm g}$. 
The data are shown together with an analytic approximation
($t_{1/2}$$\gg$$t_{g}$), which is derived in Appendix~\ref{sec:rtd} 
\begin{equation}
P_{\rm r}(t_{\rm r}) = \frac{1}{\langle\tau\rangle}
\frac{e^{-n \Theta}}{1-e^{-\Theta}} \times
\left\{
\begin{array}{lll}
\sinh \left[ \frac{t_{\rm r}}{\langle \tau \rangle} - 2(n-1) \Theta \right]
& \mbox{if} &
2(n-1)t_{1/2} < t_{\rm r} < (2n-1)t_{1/2} \\
\sinh \left[ 2n \Theta - \frac{t_{\rm r}}{\langle \tau \rangle} \right]
& \mbox{if} &
(2n-1)t_{1/2} < t_{\rm r} < 2nt_{1/2} \\
\end{array}
\right. \;.
\label{P_res1}
\end{equation}
The analytic form contains as a parameter the 
average lifetime $\langle\tau(T,H)\rangle_L$, which was measured in
independent field-reversal simulations. Thus, as for the order-parameter
distribution above, no fitting parameters are involved in the 
comparison between the simulation data and the analytical form. 
The generic feature of the rtd
is the structure of the peaks, which are centered at odd
multiples of the half-period (i.e., at $(2n-1)t_{1/2},\; n=1,2,\ldots$), 
with exponentially decaying heights, as has also been observed in simulations 
with a sinusoidally varying field \cite{ACHA99,SIDES97,SIDES98b}. 
This behavior is characteristic of systems undergoing stochastic resonance 
\cite{SR,JUNG93,GAMM95}.

In the phase diagram shown in Fig.~\ref{fig4}(a), the 
weak-noise stochastic-resonance \cite{SR}
behavior described here occurs in the wedge-shaped region between the two 
crossover curves, $T_{\rm DSP}(L)$ and $T_{\times}(L)$, 
as already discussed in Sec.~\ref{sec:MSD}. 
Analogous behavior was discussed in detail in Ref.~\cite{SIDES98b} 
for the case of a sinusoidal driving field. 

As $T$ is lowered further for the small system, the average metastable 
lifetime quickly increases, and the probability of not switching during a 
half-period, $P_{\rm not}(t_{1/2};L,T,H)$, approaches unity. As discussed in 
Sec.~\ref{sec:MSD}, the crossover temperature $T_\times(L)$ corresponds to 
$P_{\rm not} = 1/2$. Consequently, switching events become rare, and 
the central peak in $P(Q)$ essentially disappears, leaving only 
the two sharp peaks near $\pm 1$ [see Fig.~\ref{fig8}(d)]. At the same time, 
$U_L$ again becomes positive [see Fig.~\ref{fig4}(c)]. Significantly
below $T_{\times}(L)$, switching will never be observed during a finite
number of periods, and the system
is completely ``frozen'' into 
one of its two metastable wells [see Fig.~\ref{fig2}(c)]. This yields
$\langle|Q|\rangle$$\approx$$1$ and $U_L$$\approx$$2/3$, as shown
in Fig.~\ref{fig4}(b) and~(c), respectively.

\subsection{Large systems}
\label{sec:large}
For larger systems [$L$$\gtrsim$${\cal O}(10^2)$ for $t_{1/2}$=50~MCSS], 
at high temperatures the system is deeply in
the MD regime [Fig~\ref{fig4}(a)], where the lifetime is
independent of $L$. Here, $\langle\tau(T,H)\rangle$ is
significantly smaller than $t_{1/2}$. The limit cycle of the
magnetization is symmetric [Fig.~\ref{fig3}(a)], and the fluctuations in $Q$ 
are Gaussian and centered about zero [Fig.~\ref{fig9}(a) and (b)]. In this
regime, both $\langle|Q|\rangle_L$$\sim$$0$ and $U_L$$\sim$$0$, up
to finite-size effects [Fig.~\ref{fig4}(b) and (c)]. Upon lowering the
temperature, the underlying decay mode remains MD, but the lifetime
increases and eventually becomes comparable with the half-period. This
happens well before $T$ reaches $T_{\rm DSP}(L)$. 
When $\langle\tau(T,H)\rangle$ becomes approximately
equal to $t_{1/2}$, the system undergoes a genuine
continuous phase transition, the DPT \cite{SIDES99,SIDES98,KORN01}. 
The system magnetization
performs a slow ``wandering'' motion [Fig.~\ref{fig3}(b)], and the
distribution for $Q$ widens significantly [Fig.~\ref{fig9}(c)].
Below the transition, $P(Q)$ becomes bimodal [Fig.~\ref{fig9}(d)],
$\langle|Q|\rangle_L$$\sim$${\cal O}(1)$ [Fig.~\ref{fig4}(b)], and
$U_L$ approaches its ordered-phase value $2/3$ {\em without}
exhibiting negative values or a minimum [Fig.~\ref{fig4}(c)]. 
Also, $U_L$ for different large values of $L$ intersect at the temperature 
corresponding to the DPT [Figs.~\ref{fig4}(c), \ref{fig5}(c), 
\ref{fig6}(c), and~\ref{fig7}(c)], as expected for a continuous phase 
transition \cite{BIND92}. Detailed finite-size-scaling analysis of Monte 
Carlo simulations for systems that are large enough that the underlying 
metastable decay mode is MD, are found in Refs.~\cite{SIDES98,SIDES99,KORN01}.

As $T$ is reduced further, the underlying decay mode
crosses over to the SD regime at $T_{\rm DSP}(L)$. 
This leads to extremely large metastable
average lifetimes, such that $P_{\rm not}$ approaches unity. 
However, this has no effect on the observables: below the
DPT, the system is already performing asymmetric limit
cycles, confined to one of its metastable wells [Fig.\ref{fig3}(c)]. 

\subsection{Comparison}
\label{sec:comp}

From the above discussion for small and large systems, 
it is clear that the qualitative behavior
observed as the temperature is varied depends strongly on the
field amplitude, the half-period, {\em and the system size\/}. For example,
for $H$$=$$2.0J$ and $t_{1/2}$$=$$50$~MCSS one must employ $L$$\gtrsim$${\cal
O}(10^2)$ in order for the underlying metastable decay mode to become MD, 
so that the DPT is observed [Fig.~\ref{fig4}(c)]. 
For the same field amplitude and
$t_{1/2}$$=$$500$~MCSS one needs $L$$\gtrsim$${\cal O}(10^3)$ to achieve 
the same effect [Fig.~\ref{fig5}(c)]. 
For larger and larger half-periods, the
``infinite''-system DPT ($\langle\tau(T,H)\rangle$$\approx$$t_{1/2}$) 
occurs at lower and lower temperatures. However, at these low
temperatures, it takes {\em exponentially} large systems 
to be in the MD  regime, as seen from 
Eq.~(\ref{T_DSP}). By increasing $t_{1/2}$, one therefore 
quickly reaches the limit of any available computational resources.

\section{Summary and conclusion}
\label{sec:CON}

We addressed the finite-size effects of the periodic response of
spatially extended bistable systems by studying the
two-dimensional kinetic Ising model in an oscillating external
field. The intimate connection between the underlying metastable
decay modes and the periodic response of spatially extended
bistable systems has been stated and demonstrated several times
\cite{SIDES96,SIDES97,SIDES98b,SIDES99,SIDES98,KORN01}. 
On the other hand, it has
been claimed repeatedly \cite{ACHA94,CHAK99,ACHA99} that for large
enough periods (low-frequency regime), the dynamic phase
transition (DPT) becomes first-order. In the present study we focused
explicitly on demonstrating that any signatures resembling a 
first-order transition at lower temperatures 
are merely finite-size effects that disappear as $L$ is increased sufficiently. 

First, we reviewed the basics of the well-known theory of homogeneous
nucleation. Understanding the relevant time- and length scales and
the various decay modes [multi-droplet (MD) and single-droplet (SD)] 
in metastable decay,
one can estimate the important system-size dependent crossovers
for the periodic response. Next we presented extensive new simulation
results indicating that no first-order transition exists for any
frequency, and consequently, there can be no tri-critical point
separating lines of first-order and continuous dynamic
phase transitions. The behavior, correctly observed in Ref.~\cite{ACHA99} but
misinterpreted as indicating the existence of a first-order 
DPT, is due to the stochastic nature of the
underlying single-droplet metastable decay. In this regime the
system exhibits stochastic resonance. However, this behavior does 
{\em not} survive in the large-system limit with fixed field amplitude.

\acknowledgments

G.K. and M.A.N thank Z. Toroczkai for his hospitality and for
using the facilities at CNLS, Los Alamos National
Laboratory, where part of this manuscript was completed.
We acknowledge the support of NSF through Grant Nos.\ DMR-9871455, 
DMR-9981815, DMR-0120310, and DMR-0113049, and through the support of Research
Corporation Grant No.\ RI0761. 
This research used resources of the National Energy Research
Scientific Computing Center, which is supported by the Office of
Science of the U.S. Department of Energy under Contract No. DE-AC03-76SF00098.

\appendix

\section{Analytic approximation for the order-parameter distribution in the
stochastic regime}
\label{sec:pq}

In the stochastic regime, for large $t_{1/2}$, we can neglect
the growth time after the nucleation of the critical droplet and
approximate the switching process by {\em instantaneous}
switching of $m(t)$ between $\pm 1$ [Fig.~\ref{fig10}(a)]. Then, knowing that
the nucleation of a critical droplet is a Poisson process (i.e., the
probability density of the switching time is exponential)
with rate $\langle \tau \rangle^{-1}$, we can calculate
the probability density function (pdf) for the dynamic order parameter
$P(Q)$. It is important to recall that in the 
definition of $Q$ we included {\em both} types of averaging: averaging over a
period when the driving field starts with
negative value and also when it starts with a positive value in the first
half-period. Here we show the calculation of the
former case $P_{-}(Q)$. The calculation for the latter is identical and at the
end one simply has to add them together with weight $1/2$, resembling the way
the histogram was collected: $P(Q)$$=$$[P_{-}(Q)$$+$$P_{+}(Q)]/2$.

\begin{figure}[t]
\vspace*{2.0truecm}
\includegraphics{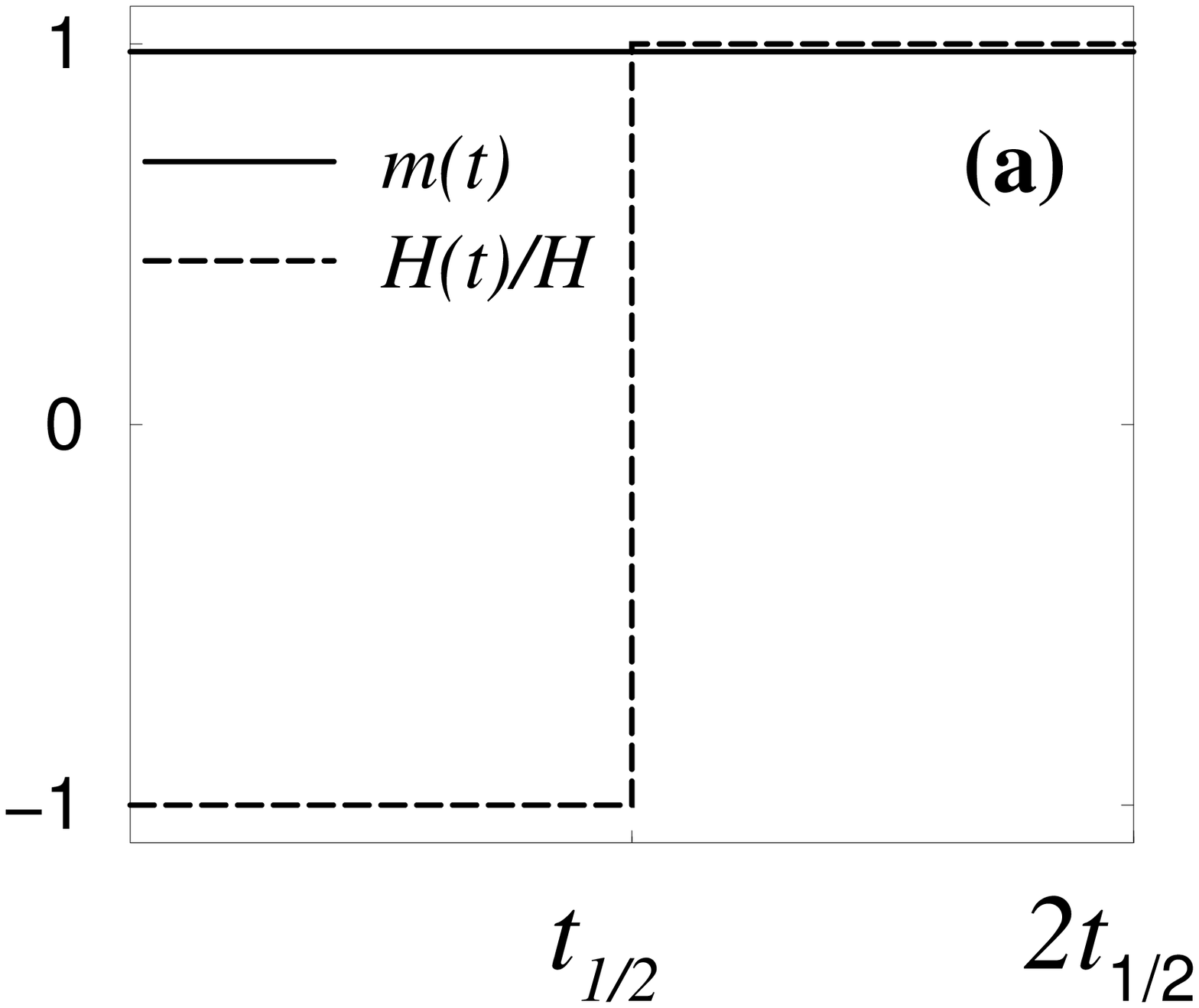}
\includegraphics{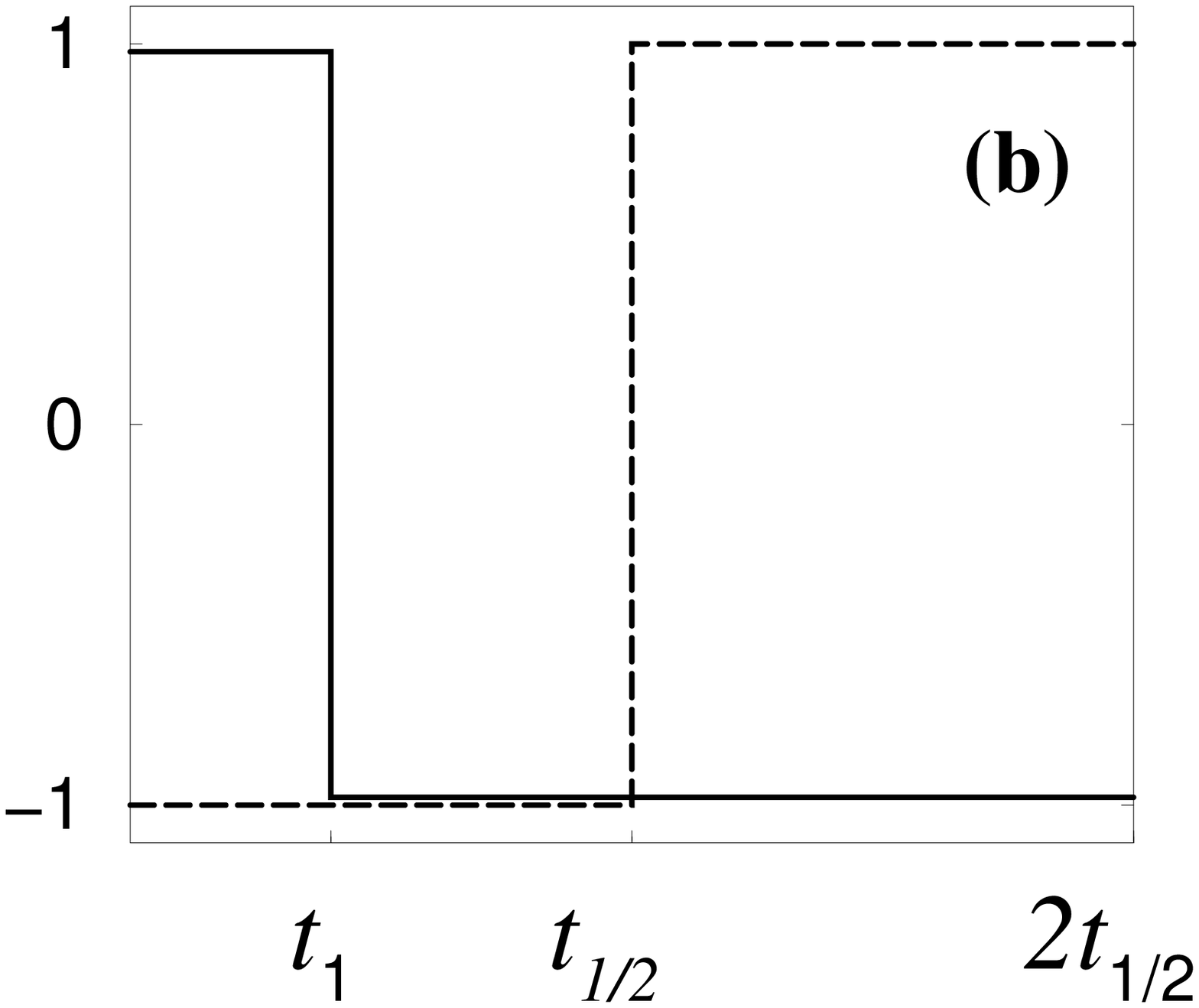}
\includegraphics{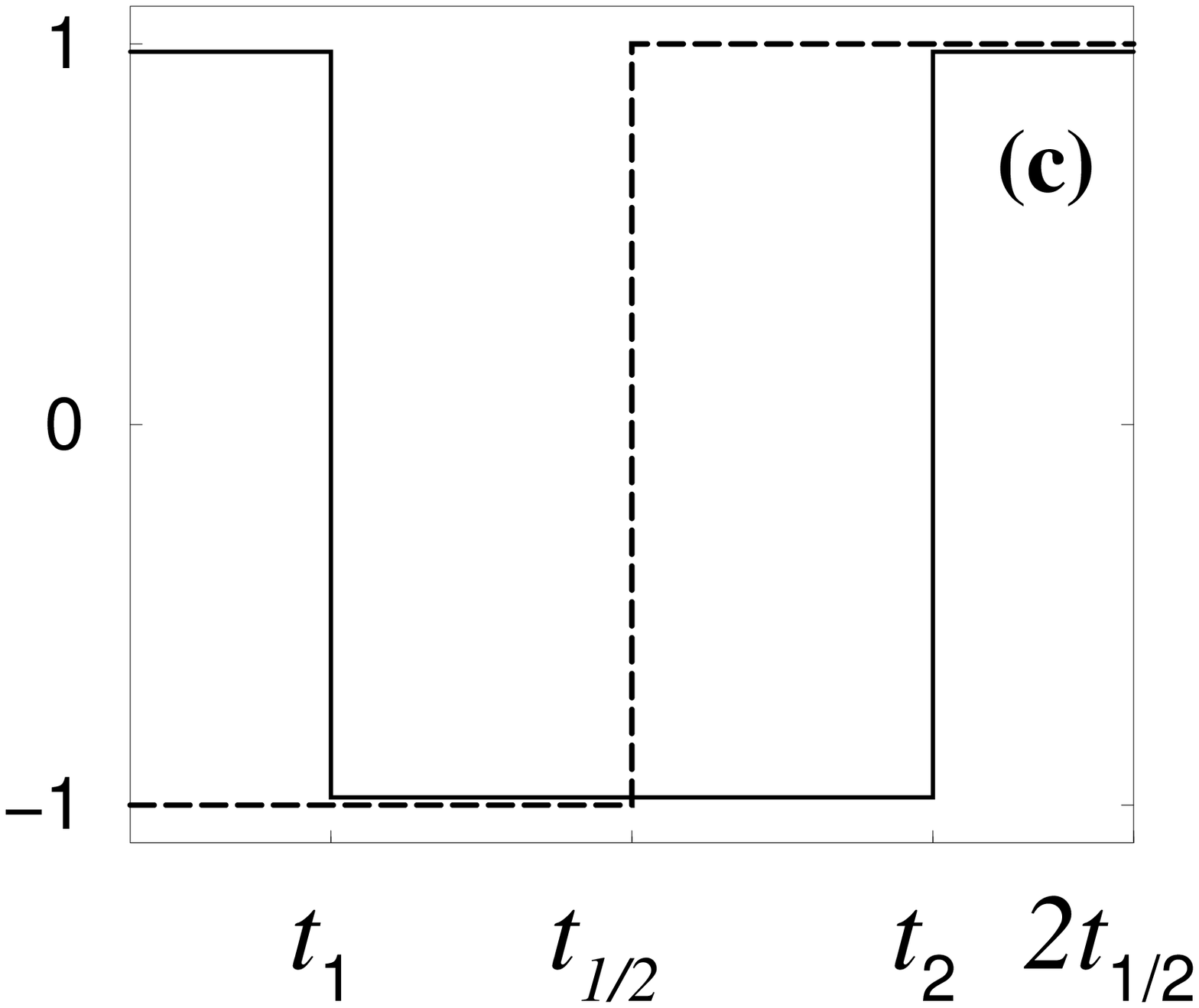}
\includegraphics{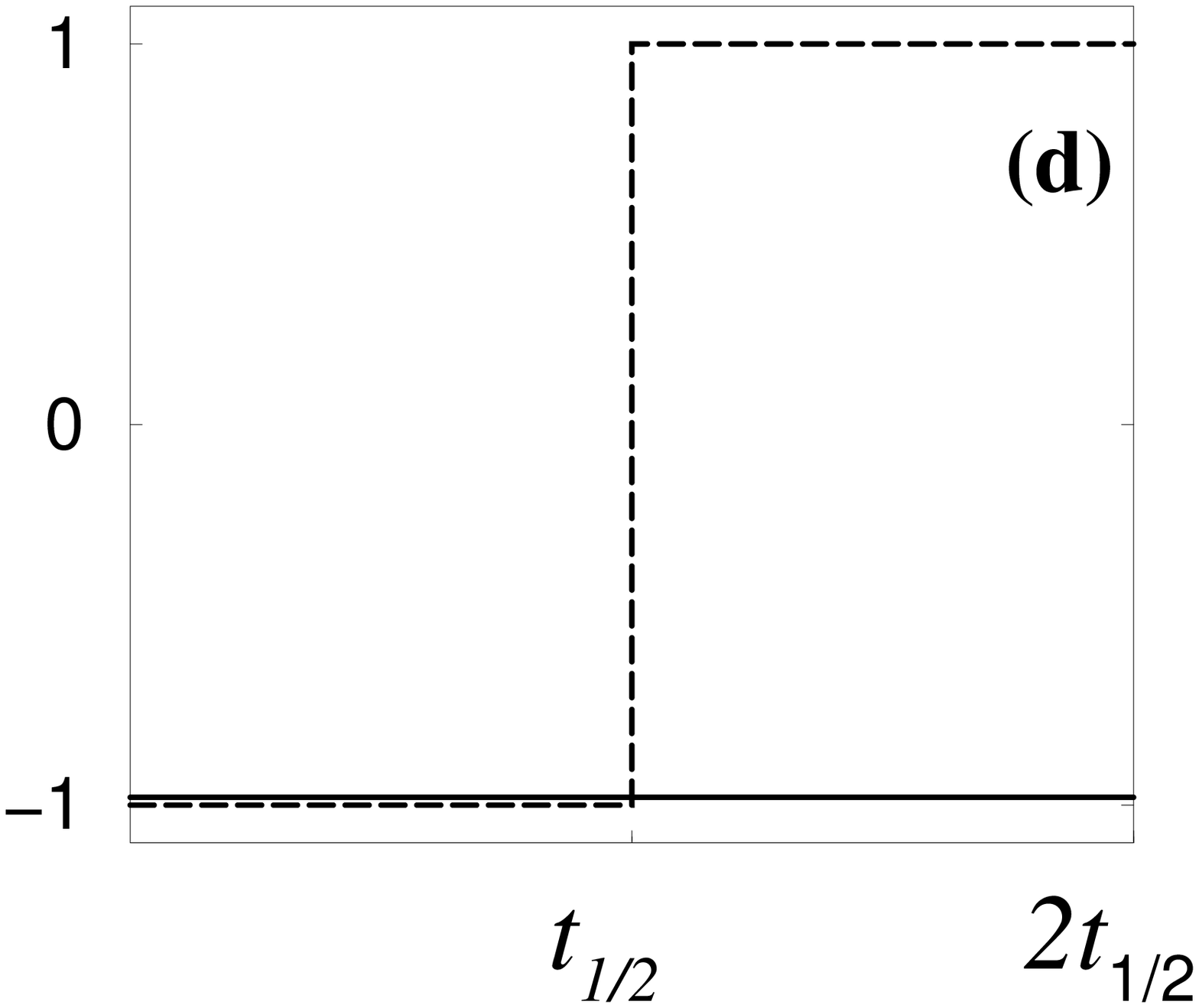}
\includegraphics{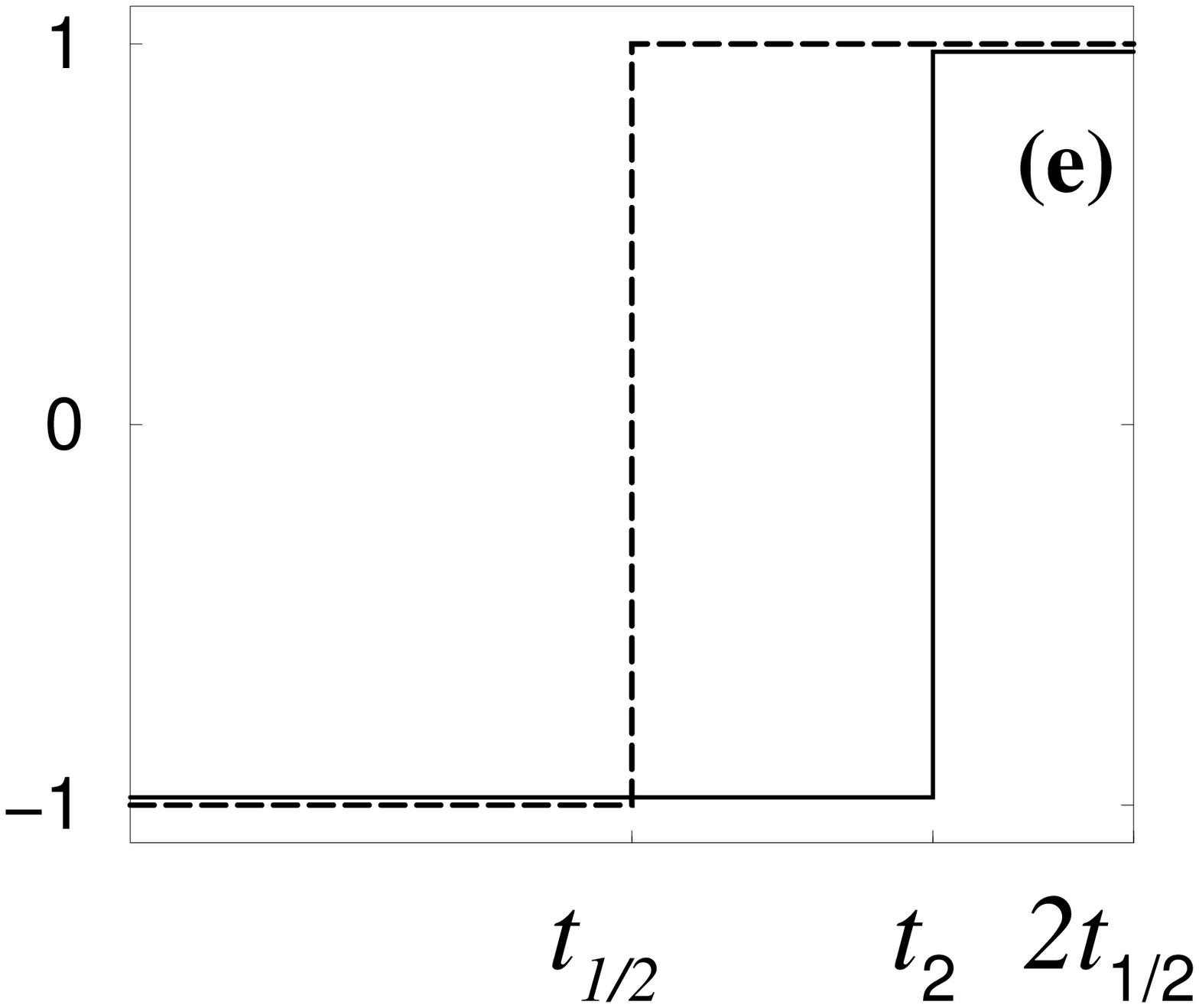}
\vspace*{2.0truecm}
\caption[]{Schematic plots for constructing the probability density $P_{-}(Q)$
when the driving field is negative in the first half-period. 
Dashed and solid lines represent the driving field and the
magnetization, respectively. The
switching times $t_1$ and $t_2$, possibly occurring in the first and
second half-period, respectively, are measured from the beginning of
their respective half-periods.}
\label{fig12}
\end{figure}

When the driving field is negative in the first half-period, there are
five mutually distinct scenarios as
illustrated on the schematic plots in Fig.~\ref{fig12}(a-e). 
In cases (a), (b), and (c) the value of the
magnetization $m(t)$ is $+1$ at the beginning of the
period, while in cases (d) and (e) it is $-1$.
Since we are interested in the {\em stationary} distribution of $Q$, first
we have to find the stationary probabilities $p^{+}_{\infty}$
($p^{-}_{\infty}$) that the 
magnetization has the value $+1$ ($-1$) at the beginning of a period.
After a quick look at the five cases above [Fig.~\ref{fig12}(a-e)], one
can write down a set of discrete-time ``time-evolution'' equations
(from one period to the next) 
\begin{eqnarray}
p^{+}_{n+1}  & = & 
\left[e^{-\Theta} + \left(1-e^{-\Theta}\right)^2\right] p^{+}_{n} +
\left(1-e^{-\Theta}\right) p^{-}_{n} \nonumber \\
p^{-}_{n+1}  & = & 
\left(1-e^{-\Theta}\right)e^{-\Theta} p^{+}_{n} +
e^{-\Theta} p^{-}_{n}
\label{pstat}
\end{eqnarray}
for $p^{+}_{n}$ and $p^{-}_{n}$, 
the probabilities that the magnetization is $+1$ and $-1$ at the beginning
of the $n$th period, respectively. In Eq.~(\ref{pstat}) we used the
definition $\Theta$$=$$t_{1/2}/\langle\tau\rangle$.
From these equations one can easily obtain the stationary-state
(fixed-point) values 
$p^{+}_{\infty}$$=$$1/(1+e^{-\Theta})$ and 
$p^{-}_{\infty}$$=$$e^{-\Theta}/(1+e^{-\Theta})$. 
Now for each case (a-e), we consider the conditional probability density of $Q$
(conditional on the value of the magnetization at the beginning 
of the period). In case (a) $Q$$=$$1$
with probability $e^{-\Theta}$, i.e., that the
magnetization does not switch in the first half-period, yielding a
delta-function contribution $e^{-\Theta}\delta(Q-1)$ 
to the full pdf. In case (b)
the magnetization switches at $t_1$ in the first
half-period and does not switch back in the second, resulting in
$Q$$=$$(t_{1}-t_{1/2})/t_{1/2}$. The contribution to the pdf is 
$e^{-\Theta}\langle\delta(Q-(t_{1}-t_{1/2})/t_{1/2})\rangle_{t_{1}}$,
where $\langle\ldots\rangle_{t_{1}}$ is an averaging over the
exponentially distributed switching time $t_{1}$. In case (c) the
magnetization switches twice, at $t_1$ in the first half-period and at
$t_2$ in the second one, resulting in
$Q$$=$$(t_{1}-t_{2})/t_{1/2}$. The contribution to the pdf is 
$\langle\delta(Q-(t_{1}-t_{2})/t_{1/2})\rangle_{t_{1},t_{2}}$, 
where $\langle\ldots\rangle_{t_{1},t_{2}}$ is an averaging over the
exponentially distributed switching times $t_{1}$ and $t_{2}$,
occurring independently in the first and second half-period (and
measured from the beginning of their respective half-periods).
In case (d) the initial value of the magnetization is $-1$ and it does
not switch in the second half-period, yielding $e^{-\Theta}\delta(Q+1)$.
In case (e) the magnetization switches once at $t_{2}$ in the second
half-period resulting in $Q$$=$$-t_{2}/t_{1/2}$ and a contribution 
$\langle\delta(Q+t_{2}/t_{1/2})\rangle_{t_{2}}$ to the pdf.
Combining the above conditional pdfs with the probabilities of the
corresponding initial values of magnetization, one obtains
\begin{eqnarray}
P_{-}(Q) & = & p^{+}_{\infty}\left\{ 
e^{-\Theta}\delta(Q-1) +
e^{-\Theta}\langle\delta(Q-(t_{1}-t_{1/2})/t_{1/2})\rangle_{t_{1}} +
\langle\delta(Q-(t_{1}-t_{2})/t_{1/2})\rangle_{t_{1},t_{2}}
\right\} \label{PQ_minus1} \\
& & + p^{-}_{\infty}\left\{
e^{-\Theta}\delta(Q+1) +
\langle\delta(Q+t_{2}/t_{1/2})\rangle_{t_{2}}
\right\} \;.\nonumber
\end{eqnarray}
Carrying out the averages above using the exponential pdfs for $t_{1}$
and $t_{2}$ we arrive at
\begin{eqnarray}
P_{-}(Q) & = & \frac{1}{1+e^{-\Theta}}\left\{ 
e^{-\Theta}\delta(Q-1) +
H(-Q)\Theta e^{-\Theta(2-|Q|)} + 
\frac{\Theta}{2}\left(
e^{-\Theta|Q|} -e^{-\Theta(2-|Q|)}
\right)
\right\} 
\label{PQ_minus2} \\
& & + \frac{e^{-\Theta}}{1+e^{-\Theta}}\left\{
e^{-\Theta}\delta(Q+1) +
H(-Q)\Theta e^{-\Theta|Q|}
\right\} \;, \nonumber
\end{eqnarray}
where $H(x)$ is the Heaviside step-function. An identical calculation
for the pdf of $Q$ for periods starting with positive value of the
driving field yields
\begin{eqnarray}
P_{+}(Q) & = & \frac{e^{-\Theta}}{1+e^{-\Theta}}\left\{ 
e^{-\Theta}\delta(Q-1) +
H(Q)\Theta e^{-\Theta|Q|}
\right\} 
\label{PQ_plus2} \\
& & + \frac{1}{1+e^{-\Theta}}\left\{
e^{-\Theta}\delta(Q+1) +
H(Q)\Theta e^{-\Theta(2-|Q|)} +
\frac{\Theta}{2}\left(
e^{-\Theta|Q|} -e^{-\Theta(2-|Q|)}
\right)
\right\} \;. \nonumber
\end{eqnarray}
The symmetrized (``phase-averaged'') dynamic order-parameter becomes
considerably simpler and easier to compare with measured histograms
\begin{equation}
P(Q) = \frac{1}{2}\left[P_{+}(Q)+P_{-}(Q)\right] =
\frac{e^{-\Theta}}{2}\delta(Q+1) +
\frac{\Theta}{2}e^{-\Theta |Q|} +
\frac{e^{-\Theta}}{2}\delta(Q-1) \;.
\label{P_Q2}
\end{equation}

\section{Residence-time distribution and its analytic approximation
in the stochastic regime}
\label{sec:rtd}

In the stochastic-resonance limit, where 
$\langle \tau \rangle$ is not much smaller than $t_{1/2}$, while both are 
much larger than $t_{\rm g}$, we can 
obtain an analytic form for the rtd $P_{\rm r}(t_{\rm r})$. 
The derivation follows that given for a sinusoidally 
oscillating field in the Appendix of 
Ref.~\cite{SIDES98b}. However, the present case is 
simpler since the probability that a switching event has not occurred 
within a time $t$ after the field has changed its sign to become
opposite to the  
magnetization direction, $P_{\rm not}(t;L,T,H)$, is a simple exponential, 
$\exp( - t / \langle \tau(T,H) \rangle_L )$ [see Eq.~(\ref{P_not})].
As a consequence, all the integrals that 
have to be evaluated numerically in the sinusoidal case, 
can here be trivially calculated analytically. 
Provided the magnetization switched in a period (say period $n$$=$$1$) at time
$t_1$ (measured from the instant the driving field changed sign), the
next magnetization switching occurring in the $n$th period at $t_2$ 
(also measured from the instant the driving field changed sign)
results in a
residence time $t_{\rm r}=(2n-1)t_{1/2}-t_{1}+t_{2}$, where $t_1$ and 
$t_2$ are exponentially distributed variables. The formal expression
for the rtd then can be written as 
\begin{equation}
P_{\rm r}(t_{\rm r}) = 
\sum^{\infty}_{n=1} \frac{e^{-(n-1)\Theta}}{1-e^{-\Theta}}
\langle\delta(t_{\rm r}-(2n-1)t_{1/2}+t_{1}-t_{2})\rangle_{t_{1},t_{2}}
\label{rtd_formal}
\end{equation}
Carrying out the averages above yields, after some rearrangement, 
\begin{equation}
P_{\rm r}(t_{\rm r}) = \frac{1}{\langle\tau\rangle}
\frac{e^{-n \Theta}}{1-e^{-\Theta}} \times
\left\{
\begin{array}{lll}
\sinh \left[ \frac{t_{\rm r}}{\langle \tau \rangle} - 2(n-1) \Theta \right]
& \mbox{if} &
2(n-1)t_{1/2} < t_{\rm r} < (2n-1)t_{1/2} \\
\sinh \left[ 2n \Theta - \frac{t_{\rm r}}{\langle \tau \rangle} \right]
& \mbox{if} &
(2n-1)t_{1/2} < t_{\rm r} < 2nt_{1/2} \\
\end{array}
\right. \;,
\label{P_res2}
\end{equation}
where $n$$=$$1,2,\ldots$.


\begin{thebibliography}{99}
\bibitem{num_values}
In our numerical solution of Eqs.~(\ref{T_DSP}) and (\ref{TX}) 
we employed the low-temperature approximation for
the classical theory of nucleation and growth where the shape-,
prefactor, and the free energy of the critical droplet are independent
of the temperature \cite{prefactor}. In particular, for the
shape factor we used $\Omega_2\approx 4$, corresponding to a square
droplet. For the the prefactor we
used $C(T,H)$$\approx$$3/8$, $1.36$, and $5/4$ for 
$H$$\in$$(J,2J)$, $H$$=$$2J$, and $H$$\in$$(2J,4J)$, respectively
\cite{prefactor}.
For the free energy of the critical droplet we used  
$F(T,H)$$\approx$$16J$$-$$6H$, $4J$, and $8J$$-$$2H$ for 
$H$$\in$$(J,2J]$, $H$$=$$2J$, and $H$$\in$$[2J,4J)$, respectively
\cite{prefactor}. The velocity $v(T,H)$ was obtained from 
Eqs.~(13) and (21) of Ref.~\cite{RIKOL00}
\end{thebibliography}
\end{document}